\documentclass[twoside,twocolumn,9pt]{article}
\usepackage{extsizes}
\usepackage[super,sort&compress,comma]{natbib} 
\usepackage[version=3]{mhchem}
\usepackage[left=1.5cm, right=1.5cm, top=1.785cm, bottom=2.0cm]{geometry}
\usepackage{balance}
\usepackage{mathptmx}
\usepackage{sectsty}
\usepackage{graphicx} 
\usepackage{lastpage}
\usepackage[format=plain,justification=justified,singlelinecheck=false,font={stretch=1.125,small,sf},labelfont=bf,labelsep=space]{caption}
\usepackage{float}
\usepackage{fancyhdr}
\usepackage{fnpos}
\usepackage[english]{babel}
\addto{\captionsenglish}{%
  
}
\usepackage{array}
\usepackage{droidsans}
\usepackage{charter}
\usepackage[T1]{fontenc}
\usepackage[usenames,dvipsnames]{xcolor}
\usepackage{setspace}
\usepackage[compact]{titlesec}
\usepackage{hyperref}

\usepackage{bm} 
\usepackage{amsmath}
\usepackage{amssymb}

\usepackage{epstopdf}

\definecolor{cream}{RGB}{222,217,201}

\begin{document}

\pagestyle{fancy}
\thispagestyle{plain}
\fancypagestyle{plain}{
\renewcommand{\headrulewidth}{0pt}
}

\makeFNbottom
\makeatletter
\renewcommand\LARGE{\@setfontsize\LARGE{15pt}{17}}
\renewcommand\Large{\@setfontsize\Large{12pt}{14}}
\renewcommand\large{\@setfontsize\large{10pt}{12}}
\renewcommand\footnotesize{\@setfontsize\footnotesize{7pt}{10}}
\makeatother

\renewcommand{\thefootnote}{\fnsymbol{footnote}}
\renewcommand\footnoterule{\vspace*{1pt}%
\color{cream}\hrule width 3.5in height 0.4pt \color{black}\vspace*{5pt}} 
\setcounter{secnumdepth}{5}

\makeatletter 
\renewcommand\@biblabel[1]{#1}            
\renewcommand\@makefntext[1]%
{\noindent\makebox[0pt][r]{\@thefnmark\,}#1}
\makeatother 
\renewcommand{\figurename}{\small{Fig.}~}
\sectionfont{\sffamily\Large}
\subsectionfont{\normalsize}
\subsubsectionfont{\bf}
\setstretch{1.125} 
\setlength{\skip\footins}{0.8cm}
\setlength{\footnotesep}{0.25cm}
\setlength{\jot}{10pt}
\titlespacing*{\section}{0pt}{4pt}{4pt}
\titlespacing*{\subsection}{0pt}{15pt}{1pt}

\fancyfoot{}
\fancyfoot[LO,RE]{\vspace{-7.1pt}\includegraphics[height=9pt]{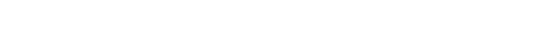}}
\fancyfoot[CO]{\vspace{-7.1pt}\hspace{13.2cm}\includegraphics{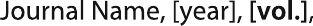}}
\fancyfoot[CE]{\vspace{-7.2pt}\hspace{-14.2cm}\includegraphics{RF}}
\fancyfoot[RO]{\footnotesize{\sffamily{1--\pageref{LastPage} ~\textbar  \hspace{2pt}\thepage}}}
\fancyfoot[LE]{\footnotesize{\sffamily{\thepage~\textbar\hspace{3.45cm} 1--\pageref{LastPage}}}}
\fancyhead{}
\renewcommand{\headrulewidth}{0pt} 
\renewcommand{\footrulewidth}{0pt}
\setlength{\arrayrulewidth}{1pt}
\setlength{\columnsep}{6.5mm}
\setlength\bibsep{1pt}

\makeatletter 
\newlength{\figrulesep} 
\setlength{\figrulesep}{0.5\textfloatsep} 

\newcommand{\topfigrule}{\vspace*{-1pt}%
\noindent{\color{cream}\rule[-\figrulesep]{\columnwidth}{1.5pt}} }

\newcommand{\botfigrule}{\vspace*{-2pt}%
\noindent{\color{cream}\rule[\figrulesep]{\columnwidth}{1.5pt}} }

\newcommand{\dblfigrule}{\vspace*{-1pt}%
\noindent{\color{cream}\rule[-\figrulesep]{\textwidth}{1.5pt}} }

\makeatother

\twocolumn[
  \begin{@twocolumnfalse}
{\includegraphics[height=30pt]{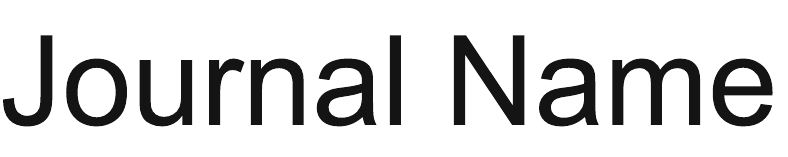}\hfill\raisebox{0pt}[0pt][0pt]{\includegraphics[height=55pt]{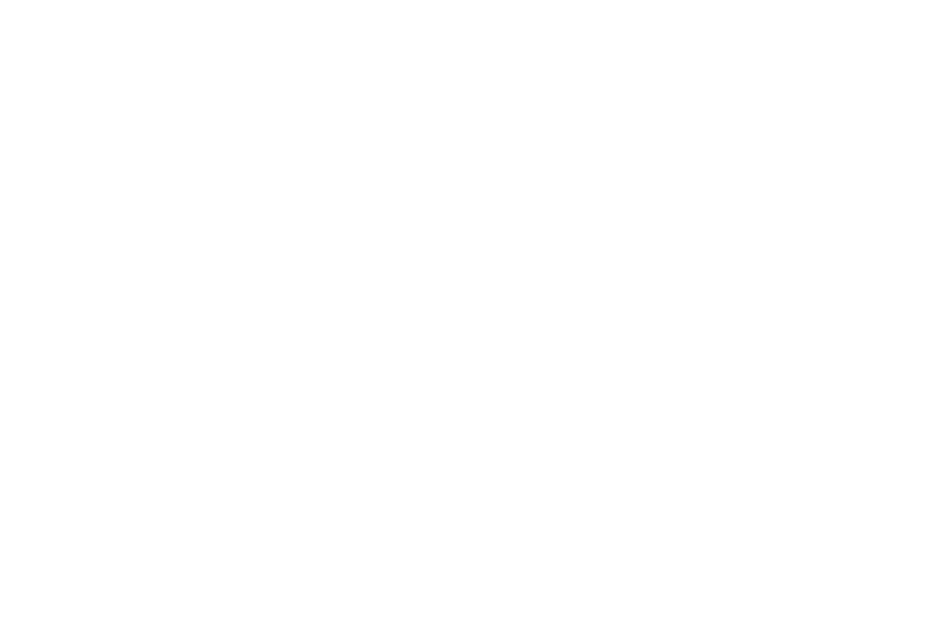}}\\[1ex]
\includegraphics[width=18.5cm]{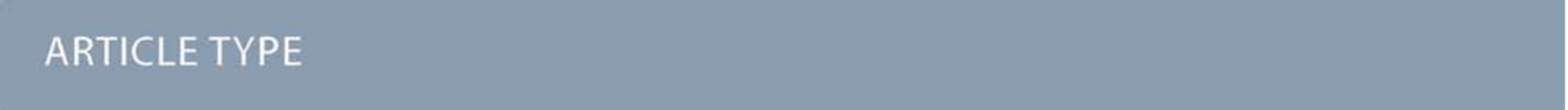}}\par
\vspace{1em}
\sffamily
\begin{tabular}{m{4.5cm} p{13.5cm} }

\includegraphics{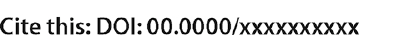} & \noindent\LARGE{\textbf{Spin transfer torques due to the bulk states of topological insulators$^\dag$}} \\
\vspace{0.3cm} & \vspace{0.3cm} \\

 & \noindent\large{James H. Cullen,$^{\ast}$\textit{$^{a}$} Rhonald Burgos Atencia,\textit{$^{a, b}$} and Dimitrie Culcer\textit{$^{a}$}} \\

\includegraphics{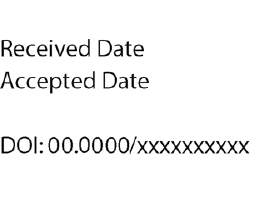} & \noindent\normalsize{Spin torques at topological insulator (TI)/ferromagnet interfaces have received considerable attention in recent years with a view towards achieving full electrical manipulation of magnetic degrees of freedom. The most important question in this field concerns the relative contributions of bulk and surface states to the spin torque, a matter that remains incompletely understood. Whereas the surface state contribution has been extensively studied, the contribution due to the bulk states has received comparatively little attention. Here we study spin torques due to TI bulk states and show that: (i) There is \textit{no spin-orbit torque} due to the bulk states on a homogeneous magnetisation, in contrast to the surface states, which give rise to a spin-orbit torque via the well-known Edelstein effect. (ii) The bulk states give rise to a \textit{spin transfer torque} (STT) due to the inhomogeneity of the magnetisation in the vicinity of the interface. This spin transfer torque, which has not been considered in TIs in the past, is somewhat unconventional since it arises from the interplay of the bulk TI spin-orbit coupling and the gradient of the monotonically decaying magnetisation inside the TI. Whereas we consider an idealised model in which the magnetisation gradient is small and the spin transfer torque is correspondingly small, we argue that in real samples the spin transfer torque should be sizable and may provide the dominant contribution due to the bulk states. We show that an experimental smoking gun for identifying the bulk states is the fact that the field-like component of the spin transfer torque generates a spin density with the same size but opposite sign for in-plane and out-of-plane magnetisations. This distinguishes them from the surface states, which are expected to give a spin density of a similar size and the same sign for both an in-plane and out-of-plane magnetisations.} \\
\end{tabular}

 \end{@twocolumnfalse} \vspace{0.6cm}

  ]

\renewcommand*\rmdefault{bch}\normalfont\upshape
\rmfamily
\section*{}
\vspace{-1cm}


\footnotetext{\textit{$^{a}$~School of Physics, The University of New South Wales, Sydney 2052, Australia.}}
\footnotetext{\textit{$^{b}$~Facultad de Ingenier\'ias, Departamento de Ciencias B\'asicas,
Universidad del Sin\'u, Cra.1{\rm w} No.38-153, 4536534,
Monter\'ia, C\'ordoba 230002, Colombia. }}

\footnotetext{\dag~Electronic Supplementary Information (ESI) available: [details of any supplementary information available should be included here]. See DOI: 00.0000/00000000.}


\section{Introduction} 

The last decade has witnessed tremendous interest in spin torques, which offer an all-electrical way to control magnetisation dynamics \cite{Nikolic_SOT, Manchon2019, brataas2012}, thereby enabling faster, more efficient operation of magnetic memory and computing devices \cite{wu2021,Shao2021RoadmapOS,Ramaswamy2018,Manchon2019}. Spin torques are especially strong in topological materials, such as topological insulators \cite{lu2022, mogi2021, che2020, Li2019, Li2014, liu2018, Zhang2018, Rodriguez-Vega2017, Song2018, Shi2018, bonell2020, wu2019, Liu2021} and Weyl and Dirac semi-metals \cite{shi2019, li2018, li2021, xie2021, ding2021, MacNeill2016}, because most of them break inversion symmetry and have strong spin-orbit coupling. The largest spin torques to date have been observed at topological insulator/ferromagnet (TI/FM) interfaces, including room-temperature magnetisation switching \cite{Wang2017, Han2017, Khang2018Acon, Dc2018}. Large spin torques have been demonstrated experimentally in a plethora of ferromagnet(ferrimagnet)/TI heterostructures, through both spin torque ferromagnetic resonance (ST-FMR)\cite{Mellnik2014,Wang2015,Jamali2015,Kondou2016,Fanchiang2018,zhu2021,Ramaswamy2019}, spin pumping\cite{Jamali2015,Baker2015,Deorani2014,Shiomi2014,Wang2016,singh2020} and harmonic Hall measurements\cite{zheng2020,Fan2022,Fan2014,fan2016}.

\begin{figure}[t!]
	\centering
	\includegraphics[width = \columnwidth]{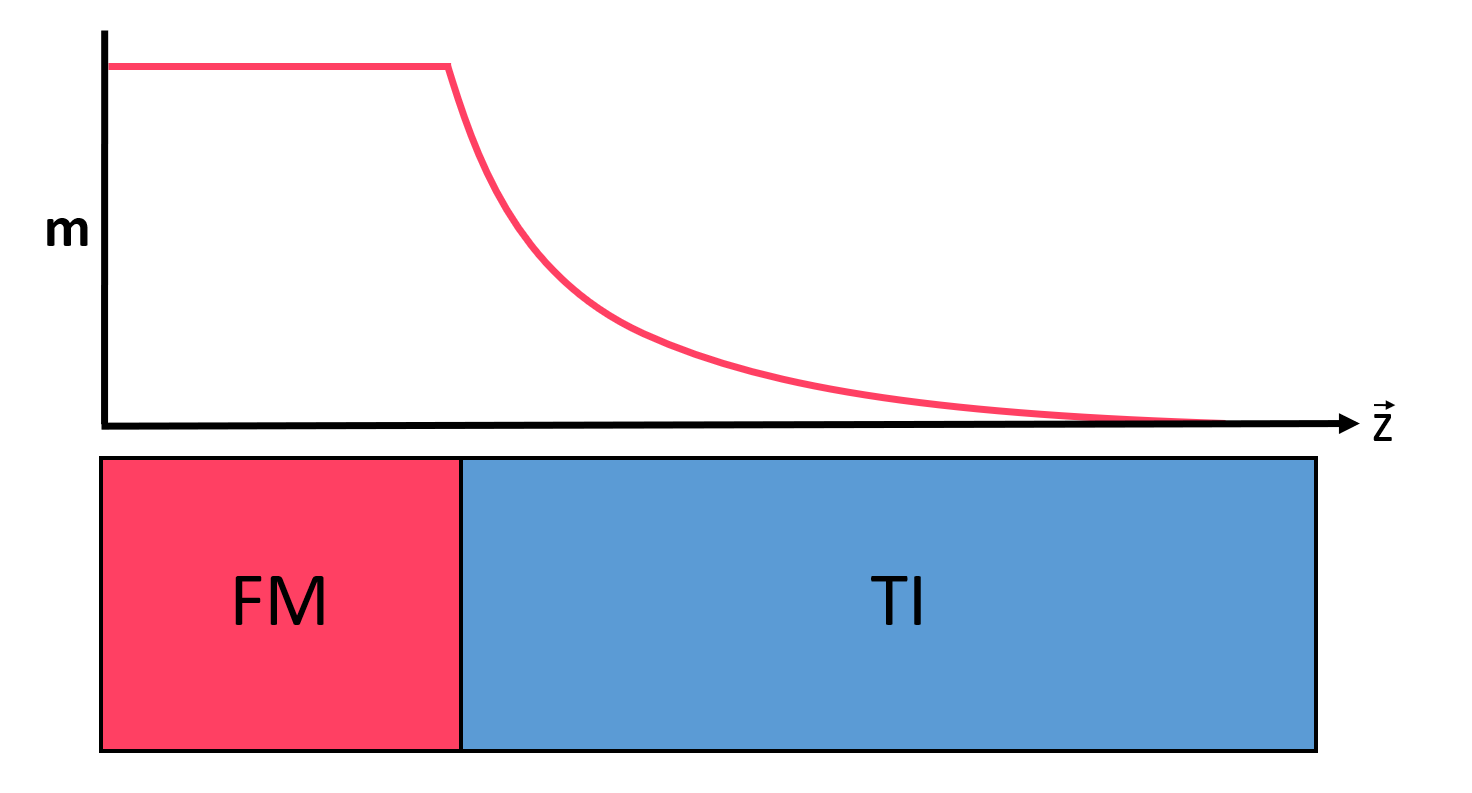}
	\caption{In our idealized model of a TI/FM heterostructure the magnetisation slowly decays into the bulk.}
\end{figure}

Topological insulator spin torques have been attributed to various mechanisms, including the Rashba-Edelstein effect (REE) in the surface states, the spin Hall effect (SHE) in the bulk \cite{Chang2015,Ghosh2018,Ado2017}, and the magnetisation penetrating a short distance into the TI \cite{Kurebayashi2019}. The extent to which each mechanism contributes is yet to be conclusively determined. The origins of the large spin torques appear to differ between experiments \cite{Jamali2015,Wang2016,gao2019,mogi2021}, which are not able to distinguish between surface and bulk contributions \cite{Wang2018}. Studies have shown that the chemical potential lies in the TI bulk conduction band for most TI/FM SOT devices \cite{Zhang2016, Marmolejo-Tejada2017}, while bulk transport dominates in a certain parameter regime\cite{Jash2021}. Hence it is believed that the bulk makes a strong contribution to the SOT, and this is customarily attributed to the spin-Hall effect \cite{Jamali2015, gao2019}, although this has never been proven. In light of this, there has been surprisingly little theoretical work on spin torques stemming from the bulk states of the TI\cite{Ghosh2018, siu2018}. Moreover, the effect of magnetisation inhomogeneity in the vicinity of the interface has never been taken into account, leaving a series of important unanswered questions: \textit{How large are bulk spin torques, what types of torques are present, and how can we distinguish bulk from surface state torques?}

\begin{figure}[t!]
	\centering
	\includegraphics[width = \columnwidth]{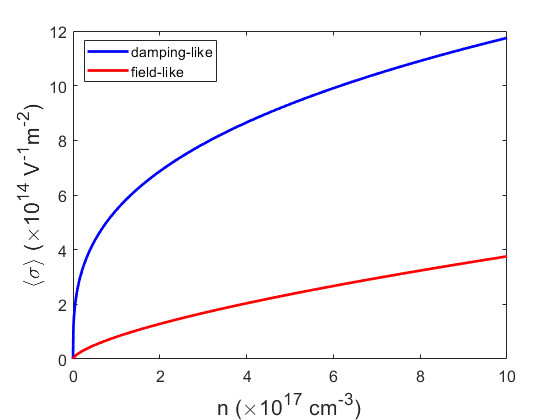}
	\caption{Dependence of the spin density per unit field on the electron number density for the magnetised Bi$_2$Se$_3$ bulk states in the weak scattering regime, with:  $|\boldsymbol{m}|=1$ meV, $\tau=1$ ps (at $n=10^{-18}$ cm$^{-3}$), $l_m=100$ nm. These results are for a magnetisation $\boldsymbol{m}\parallel\hat{x}$ or $\hat{z}$ that is decaying in the out-of-plane ($\hat{z}$) direction, while the electric field $\boldsymbol{E}\parallel\hat{x}$. The spin density shown here stems from the magnetisation gradient and is responsible for the spin transfer torque.}
\end{figure}

Motivated by these observations we develop here a quantum kinetic theory of spin torques stemming from the TI bulk and apply it to the most common TIs -- Bi$_2$Se$_3$, Bi$_2$Te$_3$ and Sb$_2$Te$_3$. We study an idealised scenario that distantly mimics the magnetic proximity effect (MPE)\cite{Vobornik2011, Eremeev2013, Lang2014, Ghosh2018} that has been demonstrated in many TI/FM insulator heterostructures\cite{Vobornik2011, Lang2014,katmis2016,lee2016,wei2013,kandala2013,mogi2021,Liu2021}. We consider a magnetised TI shown in Fig. 1. in which the magnetisation slowly decays away from the interface as $\boldsymbol{m}(\boldsymbol{r})=\boldsymbol{m}e^{-\frac{z}{l_m}}$ where $l_m k_F \gg 1$, where $k_F$ is the Fermi wave vector. The decay in real samples is much sharper, occurring over a few atomic layers \cite{Ghosh2018}, and cannot be captured in an analytical treatment, while computational methods would be prohibitively expensive. Nevertheless our analytical model captures all the essential physics of the inhomogeneous system and provides profound insight into the questions above. We show that spin dynamics in bulk TIs can be understood within the framework developed for semiconductors with an effective spin-orbit field \cite{Culcer2007, Bi2013, Culcer2017, culcer2009,winkler2008,sekine2018}, and the kinetic equation we develop captures both the weak scattering and the strong scattering (Dyakonov-Perel) regimes. Our main findings are: (i) There is no spin-orbit torque coming from the bulk, even when the states interact with a homogeneous magnetisation; (ii) For an inhomogeneous magnetisation the spin torque depends on the magnetisation gradient; its size is 1 - 2 orders of magnitude smaller than the surface state contribution for the gradients at which our theory is valid, though we argue that it may compete with the surface state contribution at the much larger gradients expected in experimental samples; (iii) For an electric field $\parallel\hat{x}$ the STT mechanism will generate a spin density $\parallel\hat{y}$, in stark contrast to the 2D case, these spins will have opposite sign for an in-plane ($\boldsymbol{m}\parallel\hat{x}$) and an out-of-plane ($\boldsymbol{m}\parallel\hat{z}$) magnetisation. This can be considered an experimental smoking gun for the bulk contribution, should it prove to be significant.

We wish to stress that we do not calculate the spin-Hall effect, mindful of complications associated with the definition of the spin current in spin-orbit coupled systems\cite{shi2006}. Our calculation is devoted entirely to the non-equilibrium spin density. The contributions independent of the magnetisation gradient essentially represent the Edelstein effect in three dimensions, while those that depend on the gradient of the magnetisation are found in the spirit of traditional spin transfer torque calculations, for example Refs.\cite{brataas2012, Tatara2004, tserkovnyak2006, Duine2007, Hals2015, hals2013}. The main idea behind this work is that the spin transfer torque, which must be present due to the magnetisation gradient, can provide a substantial contribution to the net spin torque experienced by the magnetisation, which has nothing to do with the spin-Hall effect. We believe this point has not been made previously in the field and is of crucial importance.

\begin{figure}[t!]
	\centering
	\includegraphics[width = \columnwidth]{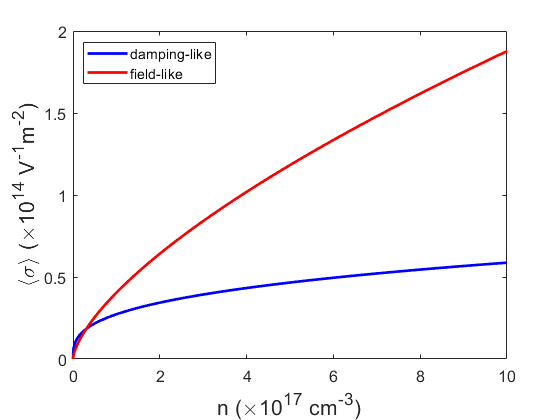}
	\caption{Dependence of the spin density per unit field on the electron  number density in the magnetised Bi$_2$Se$_3$ bulk states for:  $|\boldsymbol{m}|=1$ meV, $\tau=0.1$ ps (at $n=10^{-18}$ cm$^{-3}$), $l_m=100$ nm. These results are for a magnetisation $\boldsymbol{m}\parallel\hat{x}$ or $\hat{z}$ that is decaying in the out-of-plane ($\hat{z}$) direction with an electric field $\boldsymbol{E}\parallel\hat{x}$. The spin density shown here stems from the magnetisation gradient and is responsible for the spin transfer torque.}
\end{figure}

\section{Model and Method}
\subsection{Hamiltonian} Bulk TI states interacting with an inhomogeneous magnetisation are described by $H = \varepsilon_{\mathbf{k}}+ H_{so}$, where $\varepsilon_{\mathbf{k}} = C_{0}+C_{1} k_{z}^{2}+C_{2} k_{\parallel}^{2}$, and the spin-orbit Hamiltonian $H_{so}$ in the presence of a Zeeman field is given by \cite{Liu2010}. In the basis $\{ \frac{1}{2},-\frac{1}{2},\frac{1}{2},-\frac{1}{2} \}$ this Hamiltonian is:
\begin{equation}
	\begin{aligned}
		H_{so} & = &
		\begin{pmatrix}
			-\mathcal{M} + m_z & m_- & \mathcal{B} k_{z} & \mathcal{A}k_{-} \\
			m_+ & -\mathcal{M} - m_z & \mathcal{A} k_{+} & -\mathcal{B}k_z \\
			\mathcal{B}k_z & \mathcal{A} k_{-} & \mathcal{M} + m_z & m_- \\
			\mathcal{A} k_{+} & -\mathcal{B} k_{z} & m_+ &  \mathcal{M} - m_z
		\end{pmatrix}, 
	\end{aligned}
\end{equation}
with $\mathcal{M} = M_{0}+M_{1} k_{z}^{2}+M_{2} k_{\parallel}^{2}$, $\mathcal{A} = A_{0}+A_{2} k_{\parallel}^{2}$, $\mathcal{B} = B_{0}+B_{2} k_{z}^{2}$, $k_{\parallel}^{2} = k_{x}^{2}+k_{y}^{2}$ and $k_{\pm} = k_{x} \pm i k_{y}$. We use an effective $2\times2$ model for the conduction band, where spin-orbit coupling is represented by a wave-vector dependent effective magnetic field. The effective Hamiltonian for the conduction band using the Schrieffer-Wolff (SW) transformation\cite{Schrieffer1966,Winkler2003} is $H_{2D} = H_0 + H_Z + H_c + H_E + U$, where $H_0 = \varepsilon_{\mathbf{k}} - \mathcal{M} - \frac{\mathcal{A}^2 k_{\parallel}^2 + \mathcal{B}^2 k_{z}^2}{2\mathcal{M}}$, $H_Z = {\bm \sigma}\cdot{\bm m}$, $H_E = e{\bm E}\cdot{\bm r}$ describes the interaction with a homogeneous electric field, with ${\bm r}$ the position operator, and $U$ is the random disorder potential. The effective spin-orbit Hamiltonian is given by $H_c = \frac{\hslash}{2} \, {\bm \sigma} \cdot {\bm \Omega} \equiv \frac{\hslash}{2} \, (\sigma_z \Omega_z + \sigma_+ \Omega_- + \sigma_- \Omega_+)$, where $\sigma_\pm = (\sigma_x \pm i \sigma_y)/2$, and $\Omega_z = - \frac{\mathcal{A}^2k_{\parallel}^2}{\hslash\mathcal{M}^2} \, m_z + \frac{\mathcal{A}\mathcal{B}k_z}{\hslash\mathcal{M}^2} \, {\bm k}_{\parallel}\cdot{\bm m}_{\parallel}$, and $\Omega_\pm = \frac{\mathcal{A}\mathcal{B} k_z k_{\pm}}{\hslash\mathcal{M}^2} \, m_z - \frac{\mathcal{B}^2k_z^2}{\hslash\mathcal{M}^2}\, m_\pm + \frac{\mathcal{A}^2k_\mp}{\hslash\mathcal{M}^2} \, ({\bm k}\times{\bm m})_z$.\\

The SW transformation leading to this effective Hamiltonian is an excellent approximation due to the size of the band gap in the materials considered $\approx0.3$ eV, stemming from $M_0$, whereas the Fermi energy is of the order of a few meV. The validity of the transformation is conditional on the Fermi-surface being near the band center $k_F<3\times10^8\,m^{-1}$. We do not include hexagonal warping terms \cite{Liu2010} here. We have studied them separately and verified that they do not add any new physics. Remarkably, the spin-orbit field is entirely dependent on the magnetisation. \footnote{This has important consequences for the bulk spin Hall effect. We do not investigate this here and leave it for a future publication\cite{}. As an accurate calculation requires the full $4\times4$ Hamiltonian and the proper definition of the spin current\cite{shi2006}, which is a laborious undertaking (the conventional definition yields an unphysical spin current in insulators)\cite{Akzyanov20182, liu2015}.} \\

The SW transformation will inevitably rotate our Pauli basis. We have calculated the effect of this rotation and have confirmed that it will not significantly affect our results, this calculation is included in the supplemental material$\dag$. It simply adds some corrections that are of a higher order in the spin-orbit field and can be neglected. Furthermore, we confirmed that even when accounting for the rotated basis that the spin-orbit field vanishes when the magnetisation is 0. 

\subsection{Quantum kinetic equation} 
We use a kinetic equation formalism to calculate the linear response of the bulk states to an electric field, starting from the quantum Liouville equation as described in Refs.~\cite{Culcer2017,Bi2013}. The kinetic equation formalism that we use reproduces the results of Sinitsyn et al\cite{Sinitsyn2007} exactly as shown recently in Ref.\cite{Atencia2022}. Spin precession, which eventually leads to the spin-Hall effect, is included in our calculation, but plays only a minor role in establishing the non-equilibrium spin density. All the details of the calculation have been included in the Supplement. 

We generalize this method to incorporate an inhomogeneous magnetisation by applying a Wigner transformation \cite{Wigner1932} to the kinetic equation, which takes the form 

\begin{equation}\label{eq:kineq}
\begin{aligned}
	\frac{\partial f_E}{\partial t} + \frac{i}{\hslash}\, [H, f_E] - \frac{1}{2\hslash} \left\{ \frac{\partial H}{\partial{\bm r}} \cdot
	\frac{\partial f_E}{\partial{\bm k}} \right\} +&\\
    \frac{1}{2\hslash}\left\{ \frac{\partial H}{\partial{\bm k}} \cdot \frac{\partial f_E}{\partial{\bm r}}\right\} + \hat J (f_E) &= \frac{e{\bm E}}{\hslash}\cdot \frac{\partial f_0}{\partial{\bm k}},
\end{aligned}
\end{equation}
where  $f_0=\frac{1}{2}\left[f_{\mathrm{FD}}\left(\varepsilon_{k+}\right)+f_{\mathrm{FD}}\left(\varepsilon_{k-}\right)\right]\mathbb{I}_{2\times2} +\frac{1}{2}\left[f_{\mathrm{FD}}\left(\varepsilon_{k+}\right)-f_{\mathrm{FD}}\left(\varepsilon_{k-}\right)\right] \boldsymbol{\sigma} \cdot \boldsymbol{\hat{\Omega}_{k}}$ and $f_{\mathrm{FD}}$ is the Fermi-Dirac distribution. Without loss of generality we focus henceforth on the low temperature limit $f_{\mathrm{FD}}=\Theta(\epsilon_F-\epsilon)$. $\hat{J}$ is the scattering term in the first Born approximation$\dag$. We assume short-ranged uncorrelated spin-independent impurities.

To begin with we set the magnetisation gradient to zero, treating the system as being homogeneous and solve (3) for $f_E$. Since the system is $2\times2$ we can separate $f_E$ into $f_E=n_E\mathbb{I}_{2\times2}+\frac{1}{2}\boldsymbol{s}_E\cdot\boldsymbol{\sigma}$ in doing so we can separate the kinetic equation into a pair of coupled equations, a scalar equation $\propto \mathbb{I}$ and a spin-dependent equation $\propto \boldsymbol{\sigma}$. The spin-dependent part $\boldsymbol{s}_E$ is further separated into an angular averaged component $\overline{\boldsymbol{s}}_E$ which contains the induced spin polarisation and a remainder $\boldsymbol{t}_E$.

We find that $\overline{\boldsymbol{s}}_E=\boldsymbol{0}$, and so with a homogeneous magnetisation there is no current-induced spin polarisation. Furthermore, when there is no magnetisation the spin dependent part of the response is trivially zero $\boldsymbol{s}_E=\boldsymbol{0}$, so with no magnetisation present there is no spin polarisation and no spin currents. Next, we add the gradient terms and separate the density matrix response into homogeneous and inhomogeneous components $f_E = f_E^{hom} + f_E^{inh}$. Where $f_E^{hom}$ is the part of $f_E$ calculated by solving the kinetic equation without gradients. We then solve (3) for $f_E^{inh}$ to first order in the gradient.

These equations were solved perturbatively in the magnetisation and spin-orbit field, since the proximity-induced magnetisation is a fraction of the Fermi energy, and for the number densities in which our SW transform is valid the spin-orbit field will also be small. For our theory to be valid the magnetisation gradient must satisfy $k_F l_m \gg 1$ and for the numerical estimates we choose $|\boldsymbol{m}|=1$ meV with $l_m=100$ nm. Neither of these parameters are known for these TI/FM systems. We expect that our choice of $l_m$ is at least an order of magnitude greater than what is expected at a real interface. Hence, in the context of realistic samples, our model provides only a qualitative description of the essential physics. However, this is in keeping with the requirements for analytical theory of spin torques to be valid. No treatment, including all the theories of STT to date, can work without this assumption\cite{brataas2012, Tatara2004, tserkovnyak2006, Duine2007, Hals2015, hals2013}. Our assumption corresponds to the traditional method for calculating the STT in the vicinity of an interface\cite{Kurebayashi2019}. Whereas numerics can capture the sharp interface gradient it does not capture disorder accurately, and, in particular, doing transport numerically in the DC limit in the presence of disorder is extremely challenging due to fundamental factors, as spelled out in Ref.\cite{Culcer2017}. Hence the only pragmatic choice is to assume $l_m$ is longer that in a realistic sample for the theory to be valid, and then extrapolate to shorter lengths, which we do in the discussion section below. We stress that assuming a shorter $l_m$ changes nothing in the way our theory is formulated, it only enters the numerical estimates that follow our calculation. 

\begin{table*}[t!]
	\centering
	\begin{tabular}{ c m{0.35\columnwidth} m{0.35\columnwidth} m{0.35\columnwidth} m{0.35\columnwidth} }
		\hline
		& $m\parallel\hat{z}$, $\Omega \tau \gg 1$ & $m\parallel\hat{z}$, $\Omega \tau \ll 1$ & $m\parallel\hat{x}$, $\Omega \tau \gg 1$ & $m\parallel\hat{x}$, $\Omega \tau \ll 1$ \\
		\hline
		$\chi^{(1)}$ & $-\frac{5 e\tau^2_0 M_0^4}{3\pi^4\hslash^2 A_0^2 B_0^2 n}$ & $-\frac{5 e\tau^2_0 M_0^4}{6\pi^4\hslash^2 A_0^2 B_0^2 n}$ & $-\frac{20 e\tau^2_0 M_0^4}{3\pi^4\hslash^2 A_0^2(\Lambda^2A_0^2+B_0^2) n}$ & $-\frac{5 e\tau^2_0 M_0^4}{3\pi^4\hslash^2 A_0^2(\Lambda^2A_0^2+B_0^2) n}$ \\
		\hline
		$\chi^{(2)}$ & $\frac{e\tau_0^2 A_0 B_0 n}{20 \hslash^2 M_0^2}$ & $\frac{e\tau_0^2 A_0 B_0 n}{4 \hslash^2 M_0^2}$ & $\frac{e\tau_0^2 A_0^2 n}{20 \hslash^2 M_0^2}$ & $\frac{e\tau_0^2 A_0^2 n}{4 \hslash^2 M_0^2}$ \\
		\hline
		$\chi^{(3)}$ & $\frac{e\tau_0^2 A_0 B_0 n}{20 \hslash^2 M_0^2}$ & $\frac{e\tau_0^2 A_0 B_0 n}{4 \hslash^2 M_0^2}$ & $\frac{e\tau_0^2 A_0^2 n}{20 \hslash^2 M_0^2}$ & $\frac{e\tau_0^2 A_0^2 n}{4 \hslash^2 M_0^2}$ \\
		\hline
		$\chi^{(4)}$ & $\frac{e\tau_0 A_0 B_0 n}{40 \hslash m_z^2 M_0^2}$ & $\frac{e\tau_0 A_0 B_0 n}{8 \hslash m_z^2 M_0^2}$ & $\frac{e\tau_0 A_0^2 n}{40 \hslash m_x^2 M_0^2}$ & $\frac{e\tau_0 A_0^2 n}{8 \hslash m_x^2 M_0^2}$ \\
		\hline
		$\chi^{(5)}$ & $\frac{e\tau_0 A_0 B_0 n}{40 \hslash m_z^2 M_0^2}$ & $\frac{e\tau_0 A_0 B_0 n}{8 \hslash m_z^2 M_0^2}$ & $\frac{e\tau_0 A_0^2 n}{40 \hslash m_x^2 M_0^2}$ & $\frac{e\tau_0 A_0^2 n}{8 \hslash m_x^2 M_0^2}$ \\
		\hline
	\end{tabular}
	\caption{Values of $\chi$ for a magnetisation in the $\hat{x}$ and $\hat{z}$ direction, calculated in both the weak scattering $\Omega \tau \gg 1$ and opposite $\Omega \tau \ll 1$ limit, $\Lambda=\left[\left(C_1-M_1-\frac{B_0^2}{2M_0}\right)/\left(C_2-M_2-\frac{A_0^2}{2M_0}\right)\right]^{1/2}$ and $\tau_0=\left(2\pi^{1/3}\hslash\Lambda^{2/3}(C_2-M_2-\frac{A_0^2}{2M_0})\right)/\left(3^{1/3}n_i U_0^2 n^{1/3}\right)$}
\end{table*}

\section{Results}
Here we show the results for a fixed magnetisation in the $\hat{x}$ and $\hat{z}$ directions. For a perpendicular magnetisation $m_z$ we find the electrically induced spin polarisation
\begin{equation}
	\begin{aligned}
		\langle \boldsymbol{S}\rangle_i =& \boldsymbol{O}_{l}\chi^{(1)} (\boldsymbol{E}_l\nabla_l) m_z \hat{z}_i+\chi^{(2)} \nabla_i (\boldsymbol{E}_l m_z \hat{z}_l)+\\
		&\chi^{(3)} \boldsymbol{E}_i (\nabla_l m_z \hat{z}_l)+\chi^{(4)} \epsilon_{ijk}m_z\hat{z}_j\boldsymbol{E}_k(\nabla_l m_z \hat{z}_l)+\\
		&\chi^{(5)} \epsilon_{ijk} m_z \hat{z}_j\nabla_k(\boldsymbol{E}_l m_z \hat{z}_l)\,,
	\end{aligned}
\end{equation}
Similarly, for an in-plane magnetisation $m_x$ we find the electrically induced spin polarisation to be:
\begin{equation}
	\begin{aligned}
		\langle \boldsymbol{S}\rangle_i =& \boldsymbol{O}_{l}\chi^{(1)} (\boldsymbol{E}_l\nabla_l) m_x \hat{x}_i+\boldsymbol{X}_i\chi^{(2)} \nabla_i (\boldsymbol{E}_l m_x \hat{x}_l)+\\
		&\boldsymbol{X}_i\chi^{(3)} \boldsymbol{E}_i (\nabla_l m_x \hat{x}_l)+\boldsymbol{X}_i\chi^{(4)} \epsilon_{ijk}m_x\hat{x}_j\boldsymbol{E}_k(\nabla_l m_x \hat{x}_l)+\\
		&\boldsymbol{X}_i\chi^{(5)}\epsilon_{ijk} m_x \hat{x}_j\nabla_k(\boldsymbol{E}_l m_x \hat{x}_l)\,,
	\end{aligned}
\end{equation}
where $\boldsymbol{O}=(1,1,\Lambda^2)$ and $\boldsymbol{X}=(0,1,\frac{B_0}{A_0})$. The $\chi$ coefficients are defined in Table 1. The inhomogeneous kinetic equation was solved in both the weak scattering $\Omega_k\tau\gg1$ and DP limits $\Omega_k\tau\ll1$. Estimations based on experimental results of the 2D surface state electron mobility\cite{wang2010,choi2012,gehring2012,Wang2013} and the bulk conductivity\cite{Jash2021} indicate that Bi$_2$Se$_3$ can have a range of scattering times from $0.1$ to $2$ ps. With number densities that can vary between $0.5-50\times10^{18}$ cm$^{-3}$\cite{wang2010,choi2012, Wang2017} depending on doping, indicating that BiSe systems could be in either the weak scattering limit or DP limit. For the numerical estimates we used we chose the number density to be close to experimentally recorded values\cite{wang2010} while ensuring our approximations remain valid. We used the lower end of the range of estimated scattering times because the higher estimated scattering times came from surface state transport measurements. 

In our model the $\hat{x}$ \& $\hat{y}$ directions are equivalent so for an in-plane magnetisation $m_y$ there will be a similar electrically-induced spin polarisation to (4), one simply needs to replace $m_x$ with $m_y$ and $\hat{x}$ with $\hat{y}$. Numerical estimates indicate that for the magnetisation and decay length chosen these spin transfer torque (STT) terms are small compared to the surface state contributions to the spin torque.

The first 3 terms ($\chi^{(1)},\,\chi^{(2)}$ and $\chi^{(3)}$) are damping-like and last two terms ($\chi^{(4)},\,\chi^{(5)}$) are field-like\cite{Hals2015, hals2013}. The form of these spin polarisations is similar to the STT terms calculated for the surface states\cite{Kurebayashi2019}. For a general magnetisation where all components and gradients of $\boldsymbol{m}$ are present we found the spin polarisation to have a complex form and contains terms that have not been previously calculated for general STTs \cite{Hals2015}$\dag$. This is because the lowest order terms are fourth order in the magnetisation, which couples directly to the spin-orbit terms. 

We find that in the weak scattering limit the damping-like torque is the dominant contribution from the STT. Conversely, in the opposite scattering limit the field-like torque dominates, as seen in Fig. 2 and 3. This is primarily due to their dependence on the scattering time $\tau_0$,  as shown in Table 1, $\chi^{(1)},\,\chi^{(2)}$ and $\chi^{(3)}$ are quadratic in $\tau_0$ whereas, $\chi^{(4)}$ and $\chi^{(5)}$ are linear in $\tau_0$.

We repeated the calculations above with the hexagonal warping terms included. We found that for the number densities we are concerned with, the warping terms have a negligible effect on the induced spin polarisations. Similarly, our calculations have relied on the parabolic terms in $H_0$. This properly characterises the conduction band where our SW transform is accurate for all the materials other than Sb$_2$Te$_3$. We still expect our numerical estimates for Sb$_2$Te$_3$ to be reasonably accurate, though $k^4$ terms in the dispersion that were not considered should be included to obtain a more accurate prediction.

\begin{figure}[t!]
	\centering
	\includegraphics[width = \columnwidth]{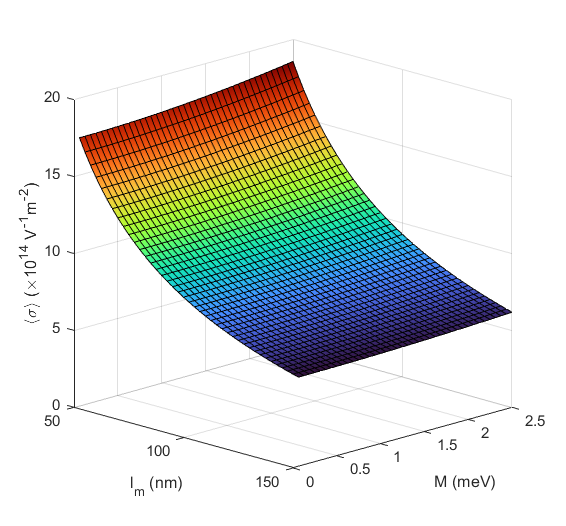}
	\caption{Total spin density per unit field vs magnetisation energy and decay length for the magnetised Bi$_2$Se$_3$ bulk states, with:  $\tau=0.1$ ps and $n=10^{-18}$ cm$^{-3}$. These results are for a magnetisation $\boldsymbol{m}\parallel\hat{x}$ or $\hat{z}$ that is decaying in the out-of-plane ($\hat{z}$) direction, while the electric field $\boldsymbol{E}\parallel\hat{x}$.}
\end{figure}

\section{Discussion} 
Our results show that the effective spin-orbit field experienced by electrons in the bulk of a TI vanishes if the magnetisation is zero. In real samples the magnetisation only penetrates a short distance into the bulk, hence TI spin transfer torques and spin-orbi torques are entirely generated by electrons near the interface. This conclusion bears some similarities to recent calculations for heavy metal spin torque devices\cite{amin2018}. Here, when the magnetisation is finite, in the vicinity of the TI/FM interface, a spin transfer torque is generated by the TI bulk states interacting with the decaying magnetisation. 
\subsection{Measuring the bulk spin transfer-torque}
We note that $m_\parallel$ experiences a spin transfer torque of the same magnitude as $m_z$. Furthermore, the field-like component of the STT spin density will have opposite sign between the cases in which the magnetisation aligned in-plane and out-of-plane as is seen in Table 2. This is an experimental smoking gun for the bulk STT contributions to the spin torque, since the REE in the topological surface states generates an in-plane polarisation\cite{sakai2014, Fischer2016, Ndiaye2017, Chang2015} that will be aligned in the same direction regardless of the magnetisation orientation. If the bulk STT provides a significant contribution to the overall spin torque, we expect the torque on an out-of-plane magnetisation will be significantly greater than the torque on an in-plane magnetisation. A simple picture of the way the STT will affect the total spin density generated in a TI/FM system is shown in Fig. 5. This analysis assumes that the prefactor of the Dirac cone surface state is positive which is consistent with Ref\cite{Liu2010}.

Consider the two systems shown in Fig. 5, for the sample with an out-of-plane magnetisation and current flowing parallel to the interface, the field-like component of the STT will be parallel to the torque generated via the REE in the surface states. The damping-like component will give a spin polarisation parallel to the applied electric field, no such spin polarisation will be generated by the surface states. So, in such a setup the bulk STT will enhance the total spin torque generated in a TI/FM system and hence increase the spin Hall angle. Conversely, for an in-plane magnetisation parallel to the applied electric field the field-like STT will suppress the surface state spin torque and hence reduce the spin Hall angle. Note, that for $\alpha<0$ this relationship would be reversed and the STT would enhance the torque on an in-plane magnetisation torque, and suppress the torque on an out-of-plane magnetisation. 

Another possible way to realise a bulk STT measurement would be by applying a electric field along the normal of the TI/FM interface and measuring the spin torque. As, in such a setup there would be a damping-like STT due to the bulk states but no contribution from the REE in the surface states. The torque comes from the $\chi^{(1)}$ terms in (3) and (4), and their expressions can be found in Table 1. For an accurate estimate of this specific torque, spin flip scattering must be included in the calculation. We leave this for a future work.

For a magnetisation aligned in-plane, the damping-like component of the STT spin density will be out-of-plane. These spins will be either parallel or anti-parallel to the out-of-plane polarisation generated due to the hexagonal warping terms in the surface states\cite{Fischer2016, Chang2015, li20191, yazyev2010} depending on the direction of magnetisation. 


\begin{table*}[t!]
	\centering
	\begin{tabular}{ c c c c c }
		\hline
		& $m\parallel\hat{x}$ &  & $m\parallel\hat{z}$  &  \\
		& $\langle\sigma_y\rangle$ /Vm$^{2}$ & $\langle\sigma_z\rangle$ /Vm$^{2}$ & $\langle\sigma_x\rangle$ /Vm$^{2}$ & $\langle\sigma_y\rangle$ /Vm$^{2}$\\
		\hline
		Bi$_2$Se$_3$ & $-1.88\times10^{14}$ & $5.87\times10^{13}$ & $5.87\times10^{13}$ & $1.88\times10^{14}$\\
		\hline
		Bi$_2$Te$_3$ & $-1.87\times10^{13}$ & $5.85\times10^{12}$ & $5.85\times10^{12}$ & $1.87\times10^{13}$ \\
		\hline
		Sb$_2$Te$_3$ & $-1.15\times10^{14}$ & $3.61\times10^{13}$ & $3.61\times10^{13}$ & $1.15\times10^{14}$\\
		\hline
	\end{tabular}
	\caption{Spin density per unit field in the magnetised bulk states for each TI evaluated for $\boldsymbol{E}\parallel\hat{x}$, $|\boldsymbol{m}|=1$ meV, $\tau=0.1$ ps, $n=10^{-18}$ cm$^{-3}$ and $l_m=100$ nm. However, here we set the magnetisation to decay in the $-\hat{z}$ direction and choose $\boldsymbol{m}(\boldsymbol{r})=\boldsymbol{m}e^{\frac{z}{l_m}}$}
\end{table*}

\begin{figure*}[t!]
	\centering
	\includegraphics[width = 2\columnwidth]{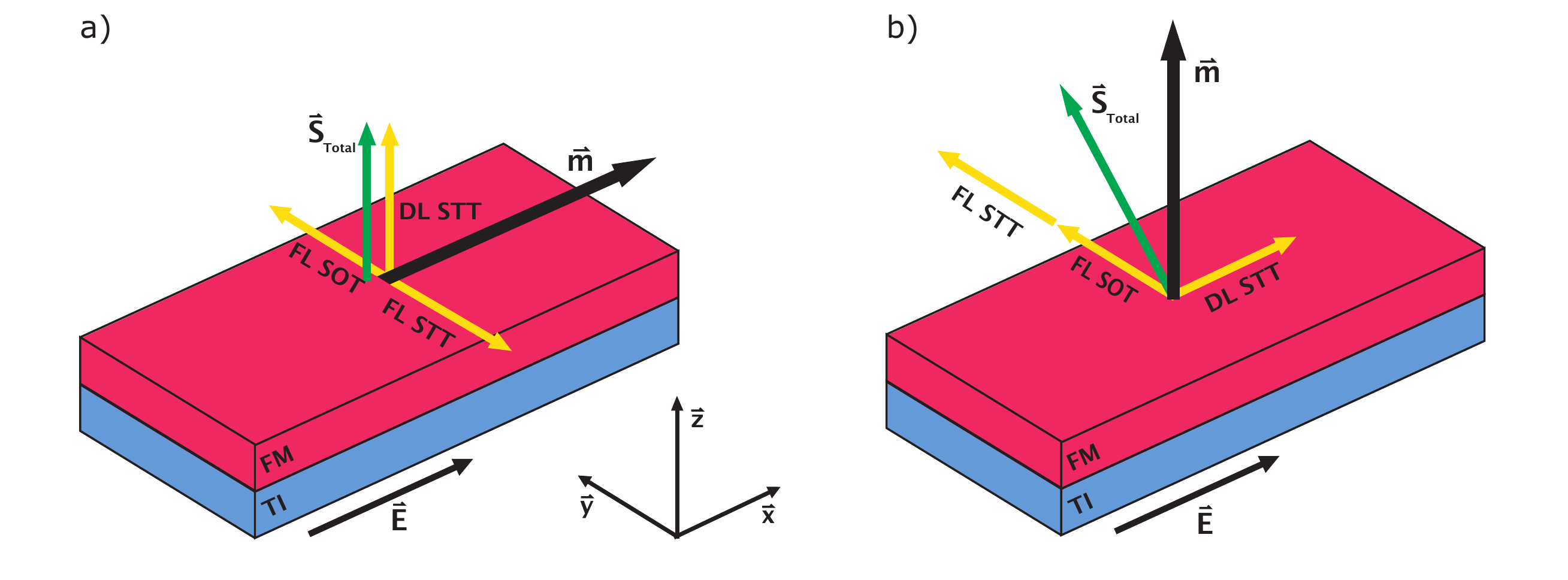}
	\caption{A simple picture demonstrating the direction of each component of the nonequillibrium spin density due to the spin transfer torque mechanism (FL and DL STT) and due to the Rashba-Edelstein effect in the surface states (FL SOT). The diagrams show the spin density generated for a system with an a) in-plane b) out-of-plane magnetisation. Here we can see that depending on the orientation of the magnetisation the field-like STT will either suppress or enhance the total spin torque in a TI/FM heterostructure, as the spins generated via the STT will be oriented either parallel or anti-parallel to the spin generated by the surface states. For the surface states we assume a simplistic picture with a Hamiltonian of the form $H=\alpha(\boldsymbol{\sigma}\times\boldsymbol{k})\cdot\boldsymbol{e}_z$ where $\alpha>0$. For a prefactor $\alpha<0$ we would see the reverse effect where the total spin torque is enhanced for an in-plane magnetisation and suppressed for an out-of-plane magnetisation.}
\end{figure*}

\subsection{Magnitude of the bulk spin transfer torque}
The 3D spin densities calculated in Table 2. can be approximately converted to 2D densities by taking them to the power of $2/3$. The spin densities calculated for Bi$_2$Se$_3$ are approximately 1 - 2 orders of magnitude smaller than the estimated 2D densities calculated at the surface of TIs\cite{Ghosh2018}. So, the bulk STT is negligible for our chosen parameters ($|\boldsymbol{m}| = 1$ meV and $l_m = 100$ nm). However, numerical calculations have found the MPE induced magnetisation in TI/FM systems to be up to an order of magnitude larger than our choice of 1 meV\cite{Eremeev2013,luo2013}, also the magnetisation decay lengths chosen for our estimates are expected to be much larger than those in real samples. In Fig. 4 we demonstrate that a smaller decay length and larger magnetisation energy can greatly increase the size of the spin density generated. Furthermore, although lying beyond the applicability of our model, one can check that, if $l_m$ is reduced to 1 nm, the bulk STT is of the same order of magnitude as the surface state SOT. This implies the possibility that the bulk STT may compete with the surface state contribution in real samples that will have much larger magnetisation gradients at the interface.

While the larger recorded number densities $n>1\times10^{18}$ cm$^{-3}$\cite{wang2010,choi2012, Wang2017} lie outside the scope of our model, we maintain that for these number densities our main conclusions remain valid, as there still cannot exist a spin-orbit field in the bulk conduction band states without a magnetisation. Numerical calculations with the full $4\times4$ Hamiltonian show that the spin-orbit field grows with $k_F$. So we expect that in these systems with larger number densities the STT will be larger as it depends on the size of the spin-orbit field. This further implies the possibility of the bulk STT competing with topological surface states spin torque.

\subsection{Magnetisation dynamics}
Here we will provide a qualitative discussion of the magnetisation dynamics of the bulk spin transfer torque. The Landau-Lifshitz-Gilbert equation is a phenomenological equation used to describe magnetisation dynamics in ferromagnetic materials. It can be used to describe spin torque dynamics in SOT devices\cite{yan2015phase, liu2012current,perez2014chiral,mikuszeit2015spin,legrand2015coherent}. This approach treats the magnetization direction $\boldsymbol{m}(r, t)$ as a classical position and time-dependent variable. The Landau-Lifshitz-Gilbert equation with the bulk TI STT terms is

\begin{equation}
    \frac{\partial \boldsymbol{m}}{\partial t} = -\gamma\boldsymbol{m}\times H_{eff} + \alpha \boldsymbol{m}\times\frac{\partial \boldsymbol{m}}{\partial t} + \frac{\gamma}{\tau_S M_S} \boldsymbol{m}\times\langle\boldsymbol{S}\rangle
\end{equation}

$H_{eff}$ contains the magnetic anisotropy, applied magnetic field and demagnetisation. $\alpha$ is the Gilbert damping coefficient, $\gamma$ is the gyromagnetic ratio, $M_S$ is the saturation magnetisation and $\tau_S$ is the spin relaxation time. The last term in this equation contains the spin torque.

Spin torques are divided into two components field-like torques and damping-like torques. Field-like torques are of the form $\tau_{FL} \sim \boldsymbol{m}\times\boldsymbol{\sigma}$, these torques cause precession in the magnetisation of the ferromagnet about $\boldsymbol{\sigma}$. Damping-like torques are of the form $\tau_{DL} \sim \boldsymbol{m}\times(\boldsymbol{\sigma}\times\boldsymbol{m})$, these torques align the magnetisation along the $\boldsymbol{\sigma}$. For the range of scatttering times we expect in TIs the spin transfer torque we calculated in topological insulators has field-like and damping-like torques of comparable magnitude. The way the combination of these two torques can manifest in magnetisation dynamics is by precession about a rotated axis. The precession is caused by the field-like torque and, the rotated axis is determined by the competition between the magnetic anisotropy of the ferromagnet and the damping-like torque. If the torque is strong enough it can cause the magnetisation to switch orientation. Both components of the torque will assist in the magnetisation switching. However, an applied external magnetic field is usually required to make the switching deterministic.
This macrospin description of spin torques does not include the Dzyaloshinskii-Moriya interaction and domain nucleation, so it is not a complete description of the physics of magnetisation switching\cite{yu2016spin,perez2014chiral,mikuszeit2015spin,legrand2015coherent}. However, for the purpose of this qualitative discussion it is sufficient.

We expect that the bulk of real TI spin torque devices will be in the strong scattering regime (DP) $\Omega_k\tau\ll1$. In sputtered samples such as the one used in Ref\cite{Dc2018} the scattering time is probably much smaller than the $0.1$ ps used in the estimates. In the STT terms calculated in this paper the field-like terms are linear in the scattering time and whereas the damping-like terms are quadratic in the scattering time. Given these scattering time dependencies we expect that the damping-like and field-like components of the STT will be either of comparable magnitude as is in Fig. 3 and Table 2 or for the field-like component to dominate. In general the spin polarisation generated from the STT mechanism differs from spin polarisation generated via the Rashba-Edelstien effect in the surface states which is purely field-like\cite{sakai2014, Fischer2016, Ndiaye2017, Chang2015}. However, this distinction is not able to be made in experiment as the diffusion of the spins into the ferromagnet tends to mix the damping-like and field-like components\cite{haney2013,manchon2012,kim2012,wang2012diff,sokolewicz2019}.

We have focused on the magnetisation gradient perpendicular to the interface because we expect it to provide the largest contribution to the STT. There will invariably be in-plane inhomogeneities in the magnetisation, for example due to interface roughness. Even though we expect these contributions to average out we have discussed results for a general magnetisation with all gradients in the Supplement$\dag$. Should the bulk TI STT prove to be significant the unconventional form of the STT terms for a general magnetisation could have important implications for the dynamics of complex spin textures such as domain walls and skyrmions in TI devices.
\subsection{Further comparisons between the STT and SOT}
The charge-to-spin conversion efficiency of the REE in the surface states is far greater than it is for the bulk STT. This is often quantified by the spin Hall angle $\theta_{\text{H}}=(\hslash/2e)\sigma_s/\sigma_{xx}$, where $\sigma_s$ is the spin conductivity and $\sigma_{xx}$ is the conductivity of the device. When the Fermi energy is in the conduction band the conductivity of the TI is an order of magnitude larger\cite{Ghosh2018} than in the insulating state where current only flows along the edges. Whereas in our numerical estimates the STT can only be up to the same order of magnitude as the surface state spin torque. So, although the STT can increase the spin conductivity, there is a greater increase in the conductivity of the device and so the spin Hall angle will be smaller in the conducting state than in the insulating state. However, this analysis may not be applicable in cases where there is significant current shunting through the FM.

We would like to note that a recent paper\cite{Ghosh2018} numerically calculated spin torques in TI/FM heterostructures with $m\parallel \hat{z}$. They modeled the TI/FM system using a tight binding approach and calculated the spin density generated in response to an applied electric field $E\parallel \hat{x}$. They found in the regime where bulk transport begins to dominate there is a crossover where the number of spins in polarised along $x$ ($S_x$) competed with $S_y$. The bulk spin Hall effect was proposed as a possible explanation. However, a numerical approach cannot explicitly discriminate spin torque mechanisms. Furthermore, in their tight binding model a steep magnetisation gradient is present. Therefore, in their numerical calculation there will be polarised spins generated via the STT mechanism. We argue that the STT calculated here could provide an alternate explanation for the bulk contribution calculated numerically for the following reasons. For lower Fermi energies where surface transport dominates the dominant spin torque mechanism will be the REE topological surface states and 2DEG Rashba states\cite{liu2018}, which will generate spins in the $\hat{y}$ and $\hat{z}$ directions. For larger Fermi energies in the conduction band and where bulk transport begins to dominate the bulk STT becomes more significant. For the system described \cite{Ghosh2018} the STT mechanism would generate a sizable $S_x$. Hence it is reasonable to assume that the STT discussed in the present work makes a significant contribution to the total spin torque seen by Ghosh and Manchon\cite{Ghosh2018}. In other words, the STT is already included within the numerics of Ghosh and Manchon \cite{Ghosh2018} and can be used to explain some of their results.

















\section{Conclusions}
In summary, we have studied electrically induced spin torques due to the bulk states of TIs in the presence of a monotonically and slowly decaying magnetisation. We have found that a homogeneous magnetisation results in no spin-orbit torque. When the magnetisation is inhomogeneous we have found a spin transfer torque, which, may compete with the surface state contribution in real samples. We also show that within our $2\times2$ model the spin-orbit field vanishes in the absence of a magnetisation. These results strongly suggest that the bulk contributions to the spin torque are almost entirely due the spin transfer torque.


\section*{Conflicts of interest}
There are no conflicts to declare.

\section*{Acknowledgements}
This project was supported by Future Fellowship FT190100062.\\
This project was supported by an Australian Government Research Training Program (RTP) Scholarship.



\balance



\begin{thebibliography}{88}%
	\makeatletter
	\providecommand \@ifxundefined [1]{%
		\@ifx{#1\undefined}
	}%
	\providecommand \@ifnum [1]{%
		\ifnum #1\expandafter \@firstoftwo
		\else \expandafter \@secondoftwo
		\fi
	}%
	\providecommand \@ifx [1]{%
		\ifx #1\expandafter \@firstoftwo
		\else \expandafter \@secondoftwo
		\fi
	}%
	\providecommand \natexlab [1]{#1}%
	\providecommand \enquote  [1]{``#1''}%
	\providecommand \bibnamefont  [1]{#1}%
	\providecommand \bibfnamefont [1]{#1}%
	\providecommand \citenamefont [1]{#1}%
	\providecommand \href@noop [0]{\@secondoftwo}%
	\providecommand \href [0]{\begingroup \@sanitize@url \@href}%
	\providecommand \@href[1]{\@@startlink{#1}\@@href}%
	\providecommand \@@href[1]{\endgroup#1\@@endlink}%
	\providecommand \@sanitize@url [0]{\catcode `\\12\catcode `\$12\catcode
		`\&12\catcode `\#12\catcode `\^12\catcode `\_12\catcode `\%12\relax}%
	\providecommand \@@startlink[1]{}%
	\providecommand \@@endlink[0]{}%
	\providecommand \url  [0]{\begingroup\@sanitize@url \@url }%
	\providecommand \@url [1]{\endgroup\@href {#1}{\urlprefix }}%
	\providecommand \urlprefix  [0]{URL }%
	\providecommand \Eprint [0]{\href }%
	\providecommand \doibase [0]{http://dx.doi.org/}%
	\providecommand \selectlanguage [0]{\@gobble}%
	\providecommand \bibinfo  [0]{\@secondoftwo}%
	\providecommand \bibfield  [0]{\@secondoftwo}%
	\providecommand \translation [1]{[#1]}%
	\providecommand \BibitemOpen [0]{}%
	\providecommand \bibitemStop [0]{}%
	\providecommand \bibitemNoStop [0]{.\EOS\space}%
	\providecommand \EOS [0]{\spacefactor3000\relax}%
	\providecommand \BibitemShut  [1]{\csname bibitem#1\endcsname}%
	\let\auto@bib@innerbib\@empty
	\bibitem [{\citenamefont {Nikolic}\ \emph {et~al.}(2018)\citenamefont
		{Nikolic}, \citenamefont {Dolui}, \citenamefont {Petrovic}, \citenamefont
		{Plechac}, \citenamefont {Markussen},\ and\ \citenamefont
		{Stokbro}}]{Nikolic_SOT}%
	\BibitemOpen
	\bibfield  {author} {\bibinfo {author} {\bibfnamefont {B.~K.}\ \bibnamefont
			{Nikoli\'c}}, \bibinfo {author} {\bibfnamefont {K.}~\bibnamefont {Dolui}},
		\bibinfo {author} {\bibfnamefont {M.~D.}\ \bibnamefont {Petrovic}}, \bibinfo
		{author} {\bibfnamefont {P.}~\bibnamefont {Plechac}}, \bibinfo {author}
		{\bibfnamefont {T.}~\bibnamefont {Markussen}}, \ and\ \bibinfo {author}
		{\bibfnamefont {K.}~\bibnamefont {Stokbro}},\ }\href@noop {} {\emph {\bibinfo
			{title} {First-Principles Quantum Transport Modeling of Spin-Transfer and
				Spin-Orbit Torques in Magnetic Multilayers}}}\ (\bibinfo  {publisher}
	{Springer},\ \bibinfo {address} {Cham},\ \bibinfo {year} {2018})\BibitemShut
	{NoStop}%
	\bibitem [{\citenamefont {Manchon}\ \emph {et~al.}(2019)\citenamefont
		{Manchon}, \citenamefont {{ˇ Zelezn}}, \citenamefont {Miron}, \citenamefont
		{Jungwirth}, \citenamefont {Sinova}, \citenamefont {Thiaville}, \citenamefont
		{Garello},\ and\ \citenamefont {Gambardella}}]{Manchon2019}%
	\BibitemOpen
	\bibfield  {author} {\bibinfo {author} {\bibfnamefont {A.}~\bibnamefont
			{Manchon}}, \bibinfo {author} {\bibfnamefont {J.}~\bibnamefont {{					Zelezn\'y}}}, \bibinfo {author} {\bibfnamefont {I.~M.}\ \bibnamefont {Miron}},
		\bibinfo {author} {\bibfnamefont {T.}~\bibnamefont {Jungwirth}}, \bibinfo
		{author} {\bibfnamefont {J.}~\bibnamefont {Sinova}}, \bibinfo {author}
		{\bibfnamefont {A.}~\bibnamefont {Thiaville}}, \bibinfo {author}
		{\bibfnamefont {K.}~\bibnamefont {Garello}}, \ and\ \bibinfo {author}
		{\bibfnamefont {P.}~\bibnamefont {Gambardella}},\ }\href@noop {} {\
		\bibinfo {pages} {arXiv 1801.09636} (\bibinfo {year} {2019})},\ \Eprint
	{http://arxiv.org/abs/1801.09636v2} {arXiv:1801.09636v2} \BibitemShut
	{NoStop}%
	\bibitem [{\citenamefont {Brataas}\ \emph {et~al.}(2012)\citenamefont
		{Brataas}, \citenamefont {Kent},\ and\ \citenamefont {Ohno}}]{brataas2012}%
	\BibitemOpen
	\bibfield  {author} {\bibinfo {author} {\bibfnamefont {A.}~\bibnamefont
			{Brataas}}, \bibinfo {author} {\bibfnamefont {A.~D.}\ \bibnamefont {Kent}}, \
		and\ \bibinfo {author} {\bibfnamefont {H.}~\bibnamefont {Ohno}},\ }\href@noop
	{} {\bibfield  {journal} {\bibinfo  {journal} {Nature materials}\ }\textbf
		{\bibinfo {volume} {11}},\ \bibinfo {pages} {372} (\bibinfo {year}
		{2012})}\BibitemShut {NoStop}%
	\bibitem [{\citenamefont {Wu}\ \emph {et~al.}(2021)\citenamefont {Wu},
		\citenamefont {Chen}, \citenamefont {Zhang}, \citenamefont {He},
		\citenamefont {Nance}, \citenamefont {Guo}, \citenamefont {Sasaki},
		\citenamefont {Shirokura}, \citenamefont {Hai}, \citenamefont {Fang} \emph
		{et~al.}}]{wu2021}%
	\BibitemOpen
	\bibfield  {author} {\bibinfo {author} {\bibfnamefont {H.}~\bibnamefont
			{Wu}}, \bibinfo {author} {\bibfnamefont {A.}~\bibnamefont {Chen}}, \bibinfo
		{author} {\bibfnamefont {P.}~\bibnamefont {Zhang}}, \bibinfo {author}
		{\bibfnamefont {H.}~\bibnamefont {He}}, \bibinfo {author} {\bibfnamefont
			{J.}~\bibnamefont {Nance}}, \bibinfo {author} {\bibfnamefont
			{C.}~\bibnamefont {Guo}}, \bibinfo {author} {\bibfnamefont {J.}~\bibnamefont
			{Sasaki}}, \bibinfo {author} {\bibfnamefont {T.}~\bibnamefont {Shirokura}},
		\bibinfo {author} {\bibfnamefont {P.~N.}\ \bibnamefont {Hai}}, \bibinfo
		{author} {\bibfnamefont {B.}~\bibnamefont {Fang}},  \emph {et~al.},\
	}\href@noop {} {\bibfield  {journal} {\bibinfo  {journal} {Nature
				communications}\ }\textbf {\bibinfo {volume} {12}},\ \bibinfo {pages} {1}
		(\bibinfo {year} {2021})}\BibitemShut {NoStop}%
	\bibitem [{\citenamefont {Shao}\ \emph {et~al.}(2021)\citenamefont {Shao},
		\citenamefont {Li}, \citenamefont {Liu}, \citenamefont {Yang}, \citenamefont
		{Fukami}, \citenamefont {Razavi}, \citenamefont {Wu}, \citenamefont {Wang},
		\citenamefont {Freimuth}, \citenamefont {Mokrousov}, \citenamefont {Stiles},
		\citenamefont {Emori}, \citenamefont {Hoffmann}, \citenamefont {Akerman},
		\citenamefont {Roy}, \citenamefont {Wang}, \citenamefont {Yang},
		\citenamefont {Garello},\ and\ \citenamefont {Zhang}}]{Shao2021RoadmapOS}%
	\BibitemOpen
	\bibfield  {author} {\bibinfo {author} {\bibfnamefont {Q.}~\bibnamefont
			{Shao}}, \bibinfo {author} {\bibfnamefont {P.}~\bibnamefont {Li}}, \bibinfo
		{author} {\bibfnamefont {L.}~\bibnamefont {Liu}}, \bibinfo {author}
		{\bibfnamefont {H.}~\bibnamefont {Yang}}, \bibinfo {author} {\bibfnamefont
			{S.}~\bibnamefont {Fukami}}, \bibinfo {author} {\bibfnamefont
			{A.}~\bibnamefont {Razavi}}, \bibinfo {author} {\bibfnamefont
			{H.}~\bibnamefont {Wu}}, \bibinfo {author} {\bibfnamefont {K.}~\bibnamefont
			{Wang}}, \bibinfo {author} {\bibfnamefont {F.}~\bibnamefont {Freimuth}},
		\bibinfo {author} {\bibfnamefont {Y.}~\bibnamefont {Mokrousov}}, \bibinfo
		{author} {\bibfnamefont {M.~D.}\ \bibnamefont {Stiles}}, \bibinfo {author}
		{\bibfnamefont {S.}~\bibnamefont {Emori}}, \bibinfo {author} {\bibfnamefont
			{A.}~\bibnamefont {Hoffmann}}, \bibinfo {author} {\bibfnamefont
			{J.}~\bibnamefont {Akerman}}, \bibinfo {author} {\bibfnamefont
			{K.}~\bibnamefont {Roy}}, \bibinfo {author} {\bibfnamefont {J.-P.}\
			\bibnamefont {Wang}}, \bibinfo {author} {\bibfnamefont {S.-H.}\ \bibnamefont
			{Yang}}, \bibinfo {author} {\bibfnamefont {K.}~\bibnamefont {Garello}}, \
		and\ \bibinfo {author} {\bibfnamefont {W.}~\bibnamefont {Zhang}},\ }\href
	{\doibase 10.1109/tmag.2021.3078583} {\bibfield  {journal} {\bibinfo
			{journal} {IEEE Transactions on Magnetics}\ }\textbf {\bibinfo {volume}
			{57}},\ \bibinfo {pages} {1–39} (\bibinfo {year} {2021})}\BibitemShut
	{NoStop}%
	\bibitem [{\citenamefont {Ramaswamy}\ \emph {et~al.}(2018)\citenamefont
		{Ramaswamy}, \citenamefont {Lee}, \citenamefont {Cai},\ and\ \citenamefont
		{Yang}}]{Ramaswamy2018}%
	\BibitemOpen
	\bibfield  {author} {\bibinfo {author} {\bibfnamefont {R.}~\bibnamefont
			{Ramaswamy}}, \bibinfo {author} {\bibfnamefont {J.~M.}\ \bibnamefont {Lee}},
		\bibinfo {author} {\bibfnamefont {K.}~\bibnamefont {Cai}}, \ and\ \bibinfo
		{author} {\bibfnamefont {H.}~\bibnamefont {Yang}},\ }\href {\doibase
		10.1063/1.5041793} {\bibfield  {journal} {\bibinfo  {journal} {Applied
				Physics Reviews}\ }\textbf {\bibinfo {volume} {5}},\ \bibinfo {pages}
		{031107} (\bibinfo {year} {2018})},\ \Eprint
	{http://arxiv.org/abs/1808.06829} {arXiv:1808.06829} \BibitemShut {NoStop}%
	\bibitem [{\citenamefont {Lu}\ \emph {et~al.}(2022)\citenamefont {Lu},
		\citenamefont {Li}, \citenamefont {Guo}, \citenamefont {Dong}, \citenamefont
		{Peng}, \citenamefont {Zha}, \citenamefont {Min}, \citenamefont {Zhou},\ and\
		\citenamefont {Liu}}]{lu2022}%
	\BibitemOpen
	\bibfield  {author} {\bibinfo {author} {\bibfnamefont {Q.}~\bibnamefont
			{Lu}}, \bibinfo {author} {\bibfnamefont {P.}~\bibnamefont {Li}}, \bibinfo
		{author} {\bibfnamefont {Z.}~\bibnamefont {Guo}}, \bibinfo {author}
		{\bibfnamefont {G.}~\bibnamefont {Dong}}, \bibinfo {author} {\bibfnamefont
			{B.}~\bibnamefont {Peng}}, \bibinfo {author} {\bibfnamefont {X.}~\bibnamefont
			{Zha}}, \bibinfo {author} {\bibfnamefont {T.}~\bibnamefont {Min}}, \bibinfo
		{author} {\bibfnamefont {Z.}~\bibnamefont {Zhou}}, \ and\ \bibinfo {author}
		{\bibfnamefont {M.}~\bibnamefont {Liu}},\ }\href@noop {} {\bibfield
		{journal} {\bibinfo  {journal} {Nature communications}\ }\textbf {\bibinfo
			{volume} {13}},\ \bibinfo {pages} {1} (\bibinfo {year} {2022})}\BibitemShut
	{NoStop}%
	\bibitem [{\citenamefont {Mogi}\ \emph {et~al.}(2021)\citenamefont {Mogi},
		\citenamefont {Yasuda}, \citenamefont {Fujimura}, \citenamefont {Yoshimi},
		\citenamefont {Ogawa}, \citenamefont {Tsukazaki}, \citenamefont {Kawamura},
		\citenamefont {Takahashi}, \citenamefont {Kawasaki},\ and\ \citenamefont
		{Tokura}}]{mogi2021}%
	\BibitemOpen
	\bibfield  {author} {\bibinfo {author} {\bibfnamefont {M.}~\bibnamefont
			{Mogi}}, \bibinfo {author} {\bibfnamefont {K.}~\bibnamefont {Yasuda}},
		\bibinfo {author} {\bibfnamefont {R.}~\bibnamefont {Fujimura}}, \bibinfo
		{author} {\bibfnamefont {R.}~\bibnamefont {Yoshimi}}, \bibinfo {author}
		{\bibfnamefont {N.}~\bibnamefont {Ogawa}}, \bibinfo {author} {\bibfnamefont
			{A.}~\bibnamefont {Tsukazaki}}, \bibinfo {author} {\bibfnamefont
			{M.}~\bibnamefont {Kawamura}}, \bibinfo {author} {\bibfnamefont {K.~S.}\
			\bibnamefont {Takahashi}}, \bibinfo {author} {\bibfnamefont {M.}~\bibnamefont
			{Kawasaki}}, \ and\ \bibinfo {author} {\bibfnamefont {Y.}~\bibnamefont
			{Tokura}},\ }\href@noop {} {\bibfield  {journal} {\bibinfo  {journal} {Nature
				communications}\ }\textbf {\bibinfo {volume} {12}},\ \bibinfo {pages} {1}
		(\bibinfo {year} {2021})}\BibitemShut {NoStop}%
	\bibitem [{\citenamefont {Che}\ \emph {et~al.}(2020)\citenamefont {Che},
		\citenamefont {Pan}, \citenamefont {Vareskic}, \citenamefont {Zou},
		\citenamefont {Pan}, \citenamefont {Zhang}, \citenamefont {Yin},
		\citenamefont {Wu}, \citenamefont {Shao}, \citenamefont {Deng} \emph
		{et~al.}}]{che2020}%
	\BibitemOpen
	\bibfield  {author} {\bibinfo {author} {\bibfnamefont {X.}~\bibnamefont
			{Che}}, \bibinfo {author} {\bibfnamefont {Q.}~\bibnamefont {Pan}}, \bibinfo
		{author} {\bibfnamefont {B.}~\bibnamefont {Vareskic}}, \bibinfo {author}
		{\bibfnamefont {J.}~\bibnamefont {Zou}}, \bibinfo {author} {\bibfnamefont
			{L.}~\bibnamefont {Pan}}, \bibinfo {author} {\bibfnamefont {P.}~\bibnamefont
			{Zhang}}, \bibinfo {author} {\bibfnamefont {G.}~\bibnamefont {Yin}}, \bibinfo
		{author} {\bibfnamefont {H.}~\bibnamefont {Wu}}, \bibinfo {author}
		{\bibfnamefont {Q.}~\bibnamefont {Shao}}, \bibinfo {author} {\bibfnamefont
			{P.}~\bibnamefont {Deng}},  \emph {et~al.},\ }\href@noop {} {\bibfield
		{journal} {\bibinfo  {journal} {Advanced Materials}\ }\textbf {\bibinfo
			{volume} {32}},\ \bibinfo {pages} {1907661} (\bibinfo {year}
		{2020})}\BibitemShut {NoStop}%
	\bibitem [{\citenamefont {Li}\ \emph {et~al.}(2019{\natexlab{a}})\citenamefont
		{Li}, \citenamefont {Kally}, \citenamefont {Zhang}, \citenamefont
		{Pillsbury}, \citenamefont {Ding}, \citenamefont {Csaba}, \citenamefont
		{Ding}, \citenamefont {Jiang}, \citenamefont {Liu}, \citenamefont {Sinclair},
		\citenamefont {Bi}, \citenamefont {DeMann}, \citenamefont {Rimal},
		\citenamefont {Zhang}, \citenamefont {Field}, \citenamefont {Tang},
		\citenamefont {Wang}, \citenamefont {Heinonen}, \citenamefont {Novosad},
		\citenamefont {Hoffmann}, \citenamefont {Samarth},\ and\ \citenamefont
		{Wu}}]{Li2019}%
	\BibitemOpen
	\bibfield  {author} {\bibinfo {author} {\bibfnamefont {P.}~\bibnamefont
			{Li}}, \bibinfo {author} {\bibfnamefont {J.}~\bibnamefont {Kally}}, \bibinfo
		{author} {\bibfnamefont {S.~S.-L.}\ \bibnamefont {Zhang}}, \bibinfo {author}
		{\bibfnamefont {T.}~\bibnamefont {Pillsbury}}, \bibinfo {author}
		{\bibfnamefont {J.}~\bibnamefont {Ding}}, \bibinfo {author} {\bibfnamefont
			{G.}~\bibnamefont {Csaba}}, \bibinfo {author} {\bibfnamefont
			{J.}~\bibnamefont {Ding}}, \bibinfo {author} {\bibfnamefont {J.~S.}\
			\bibnamefont {Jiang}}, \bibinfo {author} {\bibfnamefont {Y.}~\bibnamefont
			{Liu}}, \bibinfo {author} {\bibfnamefont {R.}~\bibnamefont {Sinclair}},
		\bibinfo {author} {\bibfnamefont {C.}~\bibnamefont {Bi}}, \bibinfo {author}
		{\bibfnamefont {A.}~\bibnamefont {DeMann}}, \bibinfo {author} {\bibfnamefont
			{G.}~\bibnamefont {Rimal}}, \bibinfo {author} {\bibfnamefont
			{W.}~\bibnamefont {Zhang}}, \bibinfo {author} {\bibfnamefont {S.~B.}\
			\bibnamefont {Field}}, \bibinfo {author} {\bibfnamefont {J.}~\bibnamefont
			{Tang}}, \bibinfo {author} {\bibfnamefont {W.}~\bibnamefont {Wang}}, \bibinfo
		{author} {\bibfnamefont {O.~G.}\ \bibnamefont {Heinonen}}, \bibinfo {author}
		{\bibfnamefont {V.}~\bibnamefont {Novosad}}, \bibinfo {author} {\bibfnamefont
			{A.}~\bibnamefont {Hoffmann}}, \bibinfo {author} {\bibfnamefont
			{N.}~\bibnamefont {Samarth}}, \ and\ \bibinfo {author} {\bibfnamefont
			{M.}~\bibnamefont {Wu}},\ }\href {\doibase 10.1126/sciadv.aaw3415} {\bibfield
		{journal} {\bibinfo  {journal} {Science Advances}\ }\textbf {\bibinfo
			{volume} {5}},\ \bibinfo {pages} {3415} (\bibinfo {year}
		{2019}{\natexlab{a}})},\ \Eprint
	{http://arxiv.org/abs/https://www.science.org/doi/pdf/10.1126/sciadv.aaw3415}
	{https://www.science.org/doi/pdf/10.1126/sciadv.aaw3415} \BibitemShut
	{NoStop}%
	\bibitem [{\citenamefont {Li}\ \emph {et~al.}(2014)\citenamefont {Li},
		\citenamefont {{Van 't Erve}}, \citenamefont {Robinson}, \citenamefont {Liu},
		\citenamefont {Li},\ and\ \citenamefont {Jonker}}]{Li2014}%
	\BibitemOpen
	\bibfield  {author} {\bibinfo {author} {\bibfnamefont {C.~H.}\ \bibnamefont
			{Li}}, \bibinfo {author} {\bibfnamefont {O.~M.~J.}\ \bibnamefont {{Van 't
					Erve}}}, \bibinfo {author} {\bibfnamefont {J.~T.}\ \bibnamefont {Robinson}},
		\bibinfo {author} {\bibfnamefont {Y.}~\bibnamefont {Liu}}, \bibinfo {author}
		{\bibfnamefont {L.}~\bibnamefont {Li}}, \ and\ \bibinfo {author}
		{\bibfnamefont {B.~T.}\ \bibnamefont {Jonker}},\ }\href {\doibase
		10.1038/NNANO.2014.16} {\bibfield  {journal} {\bibinfo  {journal} {Nature
				Nanotechnology}\ }\textbf {\bibinfo {volume} {9}},\ \bibinfo {pages} {218}
		(\bibinfo {year} {2014})}\BibitemShut {NoStop}%
	\bibitem [{\citenamefont {Liu}\ \emph {et~al.}(2018)\citenamefont {Liu},
		\citenamefont {Besbas}, \citenamefont {Wang}, \citenamefont {He},
		\citenamefont {Chen}, \citenamefont {Zhu}, \citenamefont {Wu}, \citenamefont
		{Lee}, \citenamefont {Wang}, \citenamefont {Moon} \emph {et~al.}}]{liu2018}%
	\BibitemOpen
	\bibfield  {author} {\bibinfo {author} {\bibfnamefont {Y.}~\bibnamefont
			{Liu}}, \bibinfo {author} {\bibfnamefont {J.}~\bibnamefont {Besbas}},
		\bibinfo {author} {\bibfnamefont {Y.}~\bibnamefont {Wang}}, \bibinfo {author}
		{\bibfnamefont {P.}~\bibnamefont {He}}, \bibinfo {author} {\bibfnamefont
			{M.}~\bibnamefont {Chen}}, \bibinfo {author} {\bibfnamefont {D.}~\bibnamefont
			{Zhu}}, \bibinfo {author} {\bibfnamefont {Y.}~\bibnamefont {Wu}}, \bibinfo
		{author} {\bibfnamefont {J.~M.}\ \bibnamefont {Lee}}, \bibinfo {author}
		{\bibfnamefont {L.}~\bibnamefont {Wang}}, \bibinfo {author} {\bibfnamefont
			{J.}~\bibnamefont {Moon}},  \emph {et~al.},\ }\href@noop {} {\bibfield
		{journal} {\bibinfo  {journal} {Nature communications}\ }\textbf {\bibinfo
			{volume} {9}},\ \bibinfo {pages} {1} (\bibinfo {year} {2018})}\BibitemShut
	{NoStop}%
	\bibitem [{\citenamefont {Zhang}\ \emph {et~al.}(2018)\citenamefont {Zhang},
		\citenamefont {Chan},\ and\ \citenamefont {Li}}]{Zhang2018}%
	\BibitemOpen
	\bibfield  {author} {\bibinfo {author} {\bibfnamefont {Q.}~\bibnamefont
			{Zhang}}, \bibinfo {author} {\bibfnamefont {K.~S.}\ \bibnamefont {Chan}}, \
		and\ \bibinfo {author} {\bibfnamefont {J.}~\bibnamefont {Li}},\ }\href
	{\doibase 10.1038/s41598-018-22680-4} {\bibfield  {journal} {\bibinfo
			{journal} {Scientific Reports 2018 8:1}\ }\textbf {\bibinfo {volume} {8}},\
		\bibinfo {pages} {1} (\bibinfo {year} {2018})}\BibitemShut {NoStop}%
	\bibitem [{\citenamefont {Rodriguez-Vega}\ \emph {et~al.}(2017)\citenamefont
		{Rodriguez-Vega}, \citenamefont {Schwiete}, \citenamefont {Sinova},\ and\
		\citenamefont {Rossi}}]{Rodriguez-Vega2017}%
	\BibitemOpen
	\bibfield  {author} {\bibinfo {author} {\bibfnamefont {M.}~\bibnamefont
			{Rodriguez-Vega}}, \bibinfo {author} {\bibfnamefont {G.}~\bibnamefont
			{Schwiete}}, \bibinfo {author} {\bibfnamefont {J.}~\bibnamefont {Sinova}}, \
		and\ \bibinfo {author} {\bibfnamefont {E.}~\bibnamefont {Rossi}},\ }\href
	{\doibase 10.1103/PHYSREVB.96.235419/FIGURES/9/MEDIUM} {\bibfield  {journal}
		{\bibinfo  {journal} {Physical Review B}\ }\textbf {\bibinfo {volume} {96}},\
		\bibinfo {pages} {235419} (\bibinfo {year} {2017})},\ \Eprint
	{http://arxiv.org/abs/1610.04229} {arXiv:1610.04229} \BibitemShut {NoStop}%
	\bibitem [{\citenamefont {Song}\ \emph {et~al.}(2018)\citenamefont {Song},
		\citenamefont {Soriano}, \citenamefont {Cummings}, \citenamefont {Robles},
		\citenamefont {Ordej{\'{o}}n},\ and\ \citenamefont {Roche}}]{Song2018}%
	\BibitemOpen
	\bibfield  {author} {\bibinfo {author} {\bibfnamefont {K.}~\bibnamefont
			{Song}}, \bibinfo {author} {\bibfnamefont {D.}~\bibnamefont {Soriano}},
		\bibinfo {author} {\bibfnamefont {A.~W.}\ \bibnamefont {Cummings}}, \bibinfo
		{author} {\bibfnamefont {R.}~\bibnamefont {Robles}}, \bibinfo {author}
		{\bibfnamefont {P.}~\bibnamefont {Ordej{\'{o}}n}}, \ and\ \bibinfo {author}
		{\bibfnamefont {S.}~\bibnamefont {Roche}},\ }\href {\doibase
		10.1021/ACS.NANOLETT.7B05482/SUPPL_FILE/NL7B05482_SI_001.PDF} {\bibfield
		{journal} {\bibinfo  {journal} {Nano Letters}\ }\textbf {\bibinfo {volume}
			{18}},\ \bibinfo {pages} {2033} (\bibinfo {year} {2018})},\ \Eprint
	{http://arxiv.org/abs/1806.02999} {arXiv:1806.02999} \BibitemShut {NoStop}%
	\bibitem [{\citenamefont {Shi}\ \emph {et~al.}(2018)\citenamefont {Shi},
		\citenamefont {Wang}, \citenamefont {Wang}, \citenamefont {Ramaswamy},
		\citenamefont {Shen}, \citenamefont {Moon}, \citenamefont {Zhu},
		\citenamefont {Yu}, \citenamefont {Oh}, \citenamefont {Feng}, \citenamefont {Yang}}]{Shi2018}%
	\BibitemOpen
	\bibfield  {author} {\bibinfo {author} {\bibfnamefont {S.}~\bibnamefont
			{Shi}}, \bibinfo {author} {\bibfnamefont {A.}~\bibnamefont {Wang}}, \bibinfo
		{author} {\bibfnamefont {Y.}~\bibnamefont {Wang}}, \bibinfo {author}
		{\bibfnamefont {R.}~\bibnamefont {Ramaswamy}}, \bibinfo {author}
		{\bibfnamefont {L.}~\bibnamefont {Shen}}, \bibinfo {author} {\bibfnamefont
			{J.}~\bibnamefont {Moon}}, \bibinfo {author} {\bibfnamefont {D.}~\bibnamefont
			{Zhu}}, \bibinfo {author} {\bibfnamefont {J.}~\bibnamefont {Yu}}, \bibinfo
		{author} {\bibfnamefont {S.}~\bibnamefont {Oh}}, \bibinfo {author}
		{\bibfnamefont {Y.}~\bibnamefont {Feng}}, \bibinfo {author}
		{\bibfnamefont {H.}~\bibnamefont {Yang}},\ }\href@noop {}
	{\bibfield  {journal} {\bibinfo  {journal} {Physical Review B}\ }\textbf
		{\bibinfo {volume} {97}},\ \bibinfo {pages} {041115(R)} (\bibinfo {year}
		{2018})}\BibitemShut {NoStop}%
	\bibitem [{\citenamefont {Bonell}\ \emph {et~al.}(2020)\citenamefont {Bonell},
		\citenamefont {Goto}, \citenamefont {Sauthier}, \citenamefont {Sierra},
		\citenamefont {Figueroa}, \citenamefont {Costache}, \citenamefont {Miwa},
		\citenamefont {Suzuki},\ and\ \citenamefont {Valenzuela}}]{bonell2020}%
	\BibitemOpen
	\bibfield  {author} {\bibinfo {author} {\bibfnamefont {F.}~\bibnamefont
			{Bonell}}, \bibinfo {author} {\bibfnamefont {M.}~\bibnamefont {Goto}},
		\bibinfo {author} {\bibfnamefont {G.}~\bibnamefont {Sauthier}}, \bibinfo
		{author} {\bibfnamefont {J.~F.}\ \bibnamefont {Sierra}}, \bibinfo {author}
		{\bibfnamefont {A.~I.}\ \bibnamefont {Figueroa}}, \bibinfo {author}
		{\bibfnamefont {M.~V.}\ \bibnamefont {Costache}}, \bibinfo {author}
		{\bibfnamefont {S.}~\bibnamefont {Miwa}}, \bibinfo {author} {\bibfnamefont
			{Y.}~\bibnamefont {Suzuki}}, \ and\ \bibinfo {author} {\bibfnamefont {S.~O.}\
			\bibnamefont {Valenzuela}},\ }\href@noop {} {\bibfield  {journal} {\bibinfo
			{journal} {Nano Letters}\ }\textbf {\bibinfo {volume} {20}},\ \bibinfo
		{pages} {5893} (\bibinfo {year} {2020})}\BibitemShut {NoStop}%
	\bibitem [{\citenamefont {Wu}\ \emph {et~al.}(2019)\citenamefont {Wu},
		\citenamefont {Zhang}, \citenamefont {Deng}, \citenamefont {Lan},
		\citenamefont {Pan}, \citenamefont {Razavi}, \citenamefont {Che},
		\citenamefont {Huang}, \citenamefont {Dai}, \citenamefont {Wong}, 
		\citenamefont {Han}, \citenamefont {Wong}}]{wu2019}%
	\BibitemOpen
	\bibfield  {author} {\bibinfo {author} {\bibfnamefont {H.}~\bibnamefont
			{Wu}}, \bibinfo {author} {\bibfnamefont {P.}~\bibnamefont {Zhang}}, \bibinfo
		{author} {\bibfnamefont {P.}~\bibnamefont {Deng}}, \bibinfo {author}
		{\bibfnamefont {Q.}~\bibnamefont {Lan}}, \bibinfo {author} {\bibfnamefont
			{Q.}~\bibnamefont {Pan}}, \bibinfo {author} {\bibfnamefont {S.~A.}\
			\bibnamefont {Razavi}}, \bibinfo {author} {\bibfnamefont {X.}~\bibnamefont
			{Che}}, \bibinfo {author} {\bibfnamefont {L.}~\bibnamefont {Huang}}, \bibinfo
		{author} {\bibfnamefont {B.}~\bibnamefont {Dai}}, \bibinfo {author}
		{\bibfnamefont {K.}~\bibnamefont {Wong}}, \bibinfo {author}
		{\bibfnamefont {X.}~\bibnamefont {Han}}, \bibinfo {author}
		{\bibfnamefont {K.~L.}~\bibnamefont {Wang}},\ }\href@noop {}
	{\bibfield  {journal} {\bibinfo  {journal} {Physical review letters}\
		}\textbf {\bibinfo {volume} {123}},\ \bibinfo {pages} {207205} (\bibinfo
		{year} {2019})}\BibitemShut {NoStop}%
	\bibitem [{\citenamefont {Liu}\ \emph {et~al.}(2021)\citenamefont {Liu},
		\citenamefont {Wu}, \citenamefont {Liao}, \citenamefont {Chen}, \citenamefont
		{Zhang}, \citenamefont {Xue}, \citenamefont {Yao}, \citenamefont {Song},
		\citenamefont {Wang},\ and\ \citenamefont {Kou}}]{Liu2021}%
	\BibitemOpen
	\bibfield  {author} {\bibinfo {author} {\bibfnamefont {X.}~\bibnamefont
			{Liu}}, \bibinfo {author} {\bibfnamefont {D.}~\bibnamefont {Wu}}, \bibinfo
		{author} {\bibfnamefont {L.}~\bibnamefont {Liao}}, \bibinfo {author}
		{\bibfnamefont {P.}~\bibnamefont {Chen}}, \bibinfo {author} {\bibfnamefont
			{Y.}~\bibnamefont {Zhang}}, \bibinfo {author} {\bibfnamefont
			{F.}~\bibnamefont {Xue}}, \bibinfo {author} {\bibfnamefont {Q.}~\bibnamefont
			{Yao}}, \bibinfo {author} {\bibfnamefont {C.}~\bibnamefont {Song}}, \bibinfo
		{author} {\bibfnamefont {K.~L.}\ \bibnamefont {Wang}}, \ and\ \bibinfo
		{author} {\bibfnamefont {X.}~\bibnamefont {Kou}},\ }\href@noop {} {\bibfield
		{journal} {\bibinfo  {journal} {Applied Physics Letters}\ }\textbf {\bibinfo
			{volume} {118}},\ \bibinfo {pages} {112406} (\bibinfo {year}
		{2021})}\BibitemShut {NoStop}%
	\bibitem [{\citenamefont {Shi}\ \emph {et~al.}(2019)\citenamefont {Shi},
		\citenamefont {Liang}, \citenamefont {Zhu}, \citenamefont {Cai},
		\citenamefont {Pollard}, \citenamefont {Wang}, \citenamefont {Wang},
		\citenamefont {Wang}, \citenamefont {He}, \citenamefont {Yu} \emph
		{et~al.}}]{shi2019}%
	\BibitemOpen
	\bibfield  {author} {\bibinfo {author} {\bibfnamefont {S.}~\bibnamefont
			{Shi}}, \bibinfo {author} {\bibfnamefont {S.}~\bibnamefont {Liang}}, \bibinfo
		{author} {\bibfnamefont {Z.}~\bibnamefont {Zhu}}, \bibinfo {author}
		{\bibfnamefont {K.}~\bibnamefont {Cai}}, \bibinfo {author} {\bibfnamefont
			{S.~D.}\ \bibnamefont {Pollard}}, \bibinfo {author} {\bibfnamefont
			{Y.}~\bibnamefont {Wang}}, \bibinfo {author} {\bibfnamefont {J.}~\bibnamefont
			{Wang}}, \bibinfo {author} {\bibfnamefont {Q.}~\bibnamefont {Wang}}, \bibinfo
		{author} {\bibfnamefont {P.}~\bibnamefont {He}}, \bibinfo {author}
		{\bibfnamefont {J.}~\bibnamefont {Yu}},  \emph {et~al.},\ }\href@noop {}
	{\bibfield  {journal} {\bibinfo  {journal} {Nature nanotechnology}\ }\textbf
		{\bibinfo {volume} {14}},\ \bibinfo {pages} {945} (\bibinfo {year}
		{2019})}\BibitemShut {NoStop}%
	\bibitem [{\citenamefont {Li}\ \emph {et~al.}(2018)\citenamefont {Li},
		\citenamefont {Wu}, \citenamefont {Wen}, \citenamefont {Zhang}, \citenamefont
		{Zhang}, \citenamefont {Zhang}, \citenamefont {Yu}, \citenamefont {Yang},
		\citenamefont {Manchon},\ and\ \citenamefont {xiang Zhang}}]{li2018}%
	\BibitemOpen
	\bibfield  {author} {\bibinfo {author} {\bibfnamefont {P.}~\bibnamefont
			{Li}}, \bibinfo {author} {\bibfnamefont {W.}~\bibnamefont {Wu}}, \bibinfo
		{author} {\bibfnamefont {Y.}~\bibnamefont {Wen}}, \bibinfo {author}
		{\bibfnamefont {C.}~\bibnamefont {Zhang}}, \bibinfo {author} {\bibfnamefont
			{J.}~\bibnamefont {Zhang}}, \bibinfo {author} {\bibfnamefont
			{S.}~\bibnamefont {Zhang}}, \bibinfo {author} {\bibfnamefont
			{Z.}~\bibnamefont {Yu}}, \bibinfo {author} {\bibfnamefont {S.~A.}\
			\bibnamefont {Yang}}, \bibinfo {author} {\bibfnamefont {A.}~\bibnamefont
			{Manchon}}, \ and\ \bibinfo {author} {\bibfnamefont {X.}~\bibnamefont {xiang
				Zhang}},\ }\href {\doibase 10.1038/s41467-018-06518-1} {\bibfield  {journal}
		{\bibinfo  {journal} {Nature Communications 2018 9:1}\ }\textbf {\bibinfo
			{volume} {9}},\ \bibinfo {pages} {1} (\bibinfo {year} {2018})}\BibitemShut
	{NoStop}%
	\bibitem [{\citenamefont {Li}\ \emph {et~al.}(2021)\citenamefont {Li},
		\citenamefont {Li}, \citenamefont {Hou}, \citenamefont {Mahendra},
		\citenamefont {Nien}, \citenamefont {Xue}, \citenamefont {Yi}, \citenamefont
		{Bi}, \citenamefont {Lee}, \citenamefont {Lin} \emph {et~al.}}]{li2021}%
	\BibitemOpen
	\bibfield  {author} {\bibinfo {author} {\bibfnamefont {X.}~\bibnamefont
			{Li}}, \bibinfo {author} {\bibfnamefont {P.}~\bibnamefont {Li}}, \bibinfo
		{author} {\bibfnamefont {V.~D.-H.}\ \bibnamefont {Hou}}, \bibinfo {author}
		{\bibfnamefont {D.}~\bibnamefont {Mahendra}}, \bibinfo {author}
		{\bibfnamefont {C.-H.}\ \bibnamefont {Nien}}, \bibinfo {author}
		{\bibfnamefont {F.}~\bibnamefont {Xue}}, \bibinfo {author} {\bibfnamefont
			{D.}~\bibnamefont {Yi}}, \bibinfo {author} {\bibfnamefont {C.}~\bibnamefont
			{Bi}}, \bibinfo {author} {\bibfnamefont {C.-M.}\ \bibnamefont {Lee}},
		\bibinfo {author} {\bibfnamefont {S.-J.}\ \bibnamefont {Lin}},  \emph
		{et~al.},\ }\href@noop {} {\bibfield  {journal} {\bibinfo  {journal}
			{Matter}\ }\textbf {\bibinfo {volume} {4}},\ \bibinfo {pages} {1639}
		(\bibinfo {year} {2021})}\BibitemShut {NoStop}%
	\bibitem [{\citenamefont {Xie}\ \emph {et~al.}(2021)\citenamefont {Xie},
		\citenamefont {Talapatra}, \citenamefont {Chen}, \citenamefont {Luo},\ and\
		\citenamefont {Wu}}]{xie2021}%
	\BibitemOpen
	\bibfield  {author} {\bibinfo {author} {\bibfnamefont {H.}~\bibnamefont
			{Xie}}, \bibinfo {author} {\bibfnamefont {A.}~\bibnamefont {Talapatra}},
		\bibinfo {author} {\bibfnamefont {X.}~\bibnamefont {Chen}}, \bibinfo {author}
		{\bibfnamefont {Z.}~\bibnamefont {Luo}}, \ and\ \bibinfo {author}
		{\bibfnamefont {Y.}~\bibnamefont {Wu}},\ }\href@noop {} {\bibfield  {journal}
		{\bibinfo  {journal} {Applied Physics Letters}\ }\textbf {\bibinfo {volume}
			{118}},\ \bibinfo {pages} {042401} (\bibinfo {year} {2021})}\BibitemShut
	{NoStop}%
	\bibitem [{\citenamefont {Ding}\ \emph {et~al.}(2021)\citenamefont {Ding},
		\citenamefont {Liu}, \citenamefont {Kalappattil}, \citenamefont {Zhang},
		\citenamefont {Mosendz}, \citenamefont {Erugu}, \citenamefont {Yu},
		\citenamefont {Tian}, \citenamefont {DeMann}, \citenamefont {Field} \emph
		{et~al.}}]{ding2021}%
	\BibitemOpen
	\bibfield  {author} {\bibinfo {author} {\bibfnamefont {J.}~\bibnamefont
			{Ding}}, \bibinfo {author} {\bibfnamefont {C.}~\bibnamefont {Liu}}, \bibinfo
		{author} {\bibfnamefont {V.}~\bibnamefont {Kalappattil}}, \bibinfo {author}
		{\bibfnamefont {Y.}~\bibnamefont {Zhang}}, \bibinfo {author} {\bibfnamefont
			{O.}~\bibnamefont {Mosendz}}, \bibinfo {author} {\bibfnamefont
			{U.}~\bibnamefont {Erugu}}, \bibinfo {author} {\bibfnamefont
			{R.}~\bibnamefont {Yu}}, \bibinfo {author} {\bibfnamefont {J.}~\bibnamefont
			{Tian}}, \bibinfo {author} {\bibfnamefont {A.}~\bibnamefont {DeMann}},
		\bibinfo {author} {\bibfnamefont {S.~B.}\ \bibnamefont {Field}},  \emph
		{et~al.},\ }\href@noop {} {\bibfield  {journal} {\bibinfo  {journal}
			{Advanced Materials}\ }\textbf {\bibinfo {volume} {33}},\ \bibinfo {pages}
		{2005909} (\bibinfo {year} {2021})}\BibitemShut {NoStop}%
	\bibitem [{\citenamefont {MacNeill}\ \emph {et~al.}(2016)\citenamefont
		{MacNeill}, \citenamefont {Stiehl}, \citenamefont {Guimaraes}, \citenamefont
		{Buhrman}, \citenamefont {Park},\ and\ \citenamefont {Ralph}}]{MacNeill2016}%
	\BibitemOpen
	\bibfield  {author} {\bibinfo {author} {\bibfnamefont {D.}~\bibnamefont
			{MacNeill}}, \bibinfo {author} {\bibfnamefont {G.~M.}\ \bibnamefont
			{Stiehl}}, \bibinfo {author} {\bibfnamefont {M.~H.}\ \bibnamefont
			{Guimaraes}}, \bibinfo {author} {\bibfnamefont {R.~A.}\ \bibnamefont
			{Buhrman}}, \bibinfo {author} {\bibfnamefont {J.}~\bibnamefont {Park}}, \
		and\ \bibinfo {author} {\bibfnamefont {D.~C.}\ \bibnamefont {Ralph}},\ }\href
	{\doibase 10.1038/nphys3933} {\bibfield  {journal} {\bibinfo  {journal}
			{Nature Physics 2016 13:3}\ }\textbf {\bibinfo {volume} {13}},\ \bibinfo
		{pages} {300} (\bibinfo {year} {2016})},\ \Eprint
	{http://arxiv.org/abs/1605.02712} {arXiv:1605.02712} \BibitemShut {NoStop}%
	\bibitem [{\citenamefont {Wang}\ \emph {et~al.}(2017)\citenamefont {Wang},
		\citenamefont {Zhu}, \citenamefont {Wu}, \citenamefont {Yang}, \citenamefont
		{Yu}, \citenamefont {Ramaswamy}, \citenamefont {Mishra}, \citenamefont {Shi},
		\citenamefont {Elyasi}, \citenamefont {Teo}, \citenamefont {Wu},\ and\
		\citenamefont {Yang}}]{Wang2017}%
	\BibitemOpen
	\bibfield  {author} {\bibinfo {author} {\bibfnamefont {Y.}~\bibnamefont
			{Wang}}, \bibinfo {author} {\bibfnamefont {D.}~\bibnamefont {Zhu}}, \bibinfo
		{author} {\bibfnamefont {Y.}~\bibnamefont {Wu}}, \bibinfo {author}
		{\bibfnamefont {Y.}~\bibnamefont {Yang}}, \bibinfo {author} {\bibfnamefont
			{J.}~\bibnamefont {Yu}}, \bibinfo {author} {\bibfnamefont {R.}~\bibnamefont
			{Ramaswamy}}, \bibinfo {author} {\bibfnamefont {R.}~\bibnamefont {Mishra}},
		\bibinfo {author} {\bibfnamefont {S.}~\bibnamefont {Shi}}, \bibinfo {author}
		{\bibfnamefont {M.}~\bibnamefont {Elyasi}}, \bibinfo {author} {\bibfnamefont
			{K.-L.}\ \bibnamefont {Teo}}, \bibinfo {author} {\bibfnamefont
			{Y.}~\bibnamefont {Wu}}, \ and\ \bibinfo {author} {\bibfnamefont
			{H.}~\bibnamefont {Yang}},\ }\href {\doibase 10.1038/s41467-017-01583-4}
	{\bibfield  {journal} {\bibinfo  {journal} {Nature Communications}\ }\textbf
		{\bibinfo {volume} {8}},\ \bibinfo {pages} {1364} (\bibinfo {year}
		{2017})}\BibitemShut {NoStop}%
	\bibitem [{\citenamefont {Han}\ \emph {et~al.}(2017)\citenamefont {Han},
		\citenamefont {Richardella}, \citenamefont {Siddiqui}, \citenamefont
		{Finley}, \citenamefont {Samarth},\ and\ \citenamefont {Liu}}]{Han2017}%
	\BibitemOpen
	\bibfield  {author} {\bibinfo {author} {\bibfnamefont {J.}~\bibnamefont
			{Han}}, \bibinfo {author} {\bibfnamefont {A.}~\bibnamefont {Richardella}},
		\bibinfo {author} {\bibfnamefont {S.~A.}\ \bibnamefont {Siddiqui}}, \bibinfo
		{author} {\bibfnamefont {J.}~\bibnamefont {Finley}}, \bibinfo {author}
		{\bibfnamefont {N.}~\bibnamefont {Samarth}}, \ and\ \bibinfo {author}
		{\bibfnamefont {L.}~\bibnamefont {Liu}},\ }\href {\doibase
		10.1103/PhysRevLett.119.077702} {\bibfield  {journal} {\bibinfo  {journal}
			{Physical Review Letters}\ }\textbf {\bibinfo {volume} {119}},\ \bibinfo
		{pages} {077702} (\bibinfo {year} {2017})}\BibitemShut {NoStop}%
	\bibitem [{\citenamefont {Khang}\ \emph {et~al.}(2018)\citenamefont {Khang},
		\citenamefont {Ueda},\ and\ \citenamefont {Hai}}]{Khang2018Acon}%
	\BibitemOpen
	\bibfield  {author} {\bibinfo {author} {\bibfnamefont {N.~H.~D.}\
			\bibnamefont {Khang}}, \bibinfo {author} {\bibfnamefont {Y.}~\bibnamefont
			{Ueda}}, \ and\ \bibinfo {author} {\bibfnamefont {P.~N.}\ \bibnamefont
			{Hai}},\ }\href {\doibase 10.1038/s41563-018-0137-y} {\bibfield  {journal}
		{\bibinfo  {journal} {Nat Mater}\ }\textbf {\bibinfo {volume} {17}},\
		\bibinfo {pages} {808} (\bibinfo {year} {2018})}\BibitemShut {NoStop}%
	\bibitem [{\citenamefont {Dc}\ \emph {et~al.}(2018)\citenamefont {Dc},
		\citenamefont {Grassi}, \citenamefont {Chen}, \citenamefont {Jamali},
		\citenamefont {Hickey}, \citenamefont {Zhang}, \citenamefont {Zhao},
		\citenamefont {Li}, \citenamefont {Quarterman}, \citenamefont {Lv},
		\citenamefont {Li}, \citenamefont {Manchon}, \citenamefont {Mkhoyan},
		\citenamefont {Low},\ and\ \citenamefont {Wang}}]{Dc2018}%
	\BibitemOpen
	\bibfield  {author} {\bibinfo {author} {\bibfnamefont {M.}~\bibnamefont
			{Dc}}, \bibinfo {author} {\bibfnamefont {R.}~\bibnamefont {Grassi}}, \bibinfo
		{author} {\bibfnamefont {J.-Y.}\ \bibnamefont {Chen}}, \bibinfo {author}
		{\bibfnamefont {M.}~\bibnamefont {Jamali}}, \bibinfo {author} {\bibfnamefont
			{D.~R.}\ \bibnamefont {Hickey}}, \bibinfo {author} {\bibfnamefont
			{D.}~\bibnamefont {Zhang}}, \bibinfo {author} {\bibfnamefont
			{Z.}~\bibnamefont {Zhao}}, \bibinfo {author} {\bibfnamefont {H.}~\bibnamefont
			{Li}}, \bibinfo {author} {\bibfnamefont {P.}~\bibnamefont {Quarterman}},
		\bibinfo {author} {\bibfnamefont {Y.}~\bibnamefont {Lv}}, \bibinfo {author}
		{\bibfnamefont {M.}~\bibnamefont {Li}}, \bibinfo {author} {\bibfnamefont
			{A.}~\bibnamefont {Manchon}}, \bibinfo {author} {\bibfnamefont {K.~A.}\
			\bibnamefont {Mkhoyan}}, \bibinfo {author} {\bibfnamefont {T.}~\bibnamefont
			{Low}}, \ and\ \bibinfo {author} {\bibfnamefont {J.-P.}\ \bibnamefont
			{Wang}},\ }\href {\doibase 10.1038/s41563-018-0136-z} {\bibfield  {journal}
		{\bibinfo  {journal} {Nature Materials}\ }\textbf {\bibinfo {volume} {17}},\
		\bibinfo {pages} {800} (\bibinfo {year} {2018})}\BibitemShut {NoStop}%
	\bibitem [{\citenamefont {Mellnik}\ \emph {et~al.}(2014)\citenamefont
		{Mellnik}, \citenamefont {Lee}, \citenamefont {Richardella}, \citenamefont
		{Grab}, \citenamefont {Mintun}, \citenamefont {Fischer}, \citenamefont
		{Vaezi}, \citenamefont {Manchon}, \citenamefont {Kim}, \citenamefont
		{Samarth},\ and\ \citenamefont {Ralph}}]{Mellnik2014}%
	\BibitemOpen
	\bibfield  {author} {\bibinfo {author} {\bibfnamefont {A.~R.}\ \bibnamefont
			{Mellnik}}, \bibinfo {author} {\bibfnamefont {J.~S.}\ \bibnamefont {Lee}},
		\bibinfo {author} {\bibfnamefont {A.}~\bibnamefont {Richardella}}, \bibinfo
		{author} {\bibfnamefont {J.~L.}\ \bibnamefont {Grab}}, \bibinfo {author}
		{\bibfnamefont {P.~J.}\ \bibnamefont {Mintun}}, \bibinfo {author}
		{\bibfnamefont {M.~H.}\ \bibnamefont {Fischer}}, \bibinfo {author}
		{\bibfnamefont {A.}~\bibnamefont {Vaezi}}, \bibinfo {author} {\bibfnamefont
			{A.}~\bibnamefont {Manchon}}, \bibinfo {author} {\bibfnamefont {E.-A.}\
			\bibnamefont {Kim}}, \bibinfo {author} {\bibfnamefont {N.}~\bibnamefont
			{Samarth}}, \ and\ \bibinfo {author} {\bibfnamefont {D.~C.}\ \bibnamefont
			{Ralph}},\ }\href {\doibase 10.1038/nature13534} {\bibfield  {journal}
		{\bibinfo  {journal} {Nature}\ }\textbf {\bibinfo {volume} {511}},\ \bibinfo
		{pages} {449–451} (\bibinfo {year} {2014})}\BibitemShut {NoStop}%
	\bibitem [{\citenamefont {Wang}\ \emph {et~al.}(2015)\citenamefont {Wang},
		\citenamefont {Deorani}, \citenamefont {Banerjee}, \citenamefont {Koirala},
		\citenamefont {Brahlek}, \citenamefont {Oh},\ and\ \citenamefont
		{Yang}}]{Wang2015}%
	\BibitemOpen
	\bibfield  {author} {\bibinfo {author} {\bibfnamefont {Y.}~\bibnamefont
			{Wang}}, \bibinfo {author} {\bibfnamefont {P.}~\bibnamefont {Deorani}},
		\bibinfo {author} {\bibfnamefont {K.}~\bibnamefont {Banerjee}}, \bibinfo
		{author} {\bibfnamefont {N.}~\bibnamefont {Koirala}}, \bibinfo {author}
		{\bibfnamefont {M.}~\bibnamefont {Brahlek}}, \bibinfo {author} {\bibfnamefont
			{S.}~\bibnamefont {Oh}}, \ and\ \bibinfo {author} {\bibfnamefont
			{H.}~\bibnamefont {Yang}},\ }\href {\doibase 10.1103/PhysRevLett.114.257202}
	{\bibfield  {journal} {\bibinfo  {journal} {Physical Review Letters}\
		}\textbf {\bibinfo {volume} {114}},\ \bibinfo {pages} {257202} (\bibinfo
		{year} {2015})}\BibitemShut {NoStop}%
	\bibitem [{\citenamefont {Jamali}\ \emph {et~al.}(2015)\citenamefont {Jamali},
		\citenamefont {Lee}, \citenamefont {Jeong}, \citenamefont {Mahfouzi},
		\citenamefont {Lv}, \citenamefont {Zhao}, \citenamefont {Nikoli\'c},
		\citenamefont {Mkhoyan}, \citenamefont {Samarth},\ and\ \citenamefont
		{Wang}}]{Jamali2015}%
	\BibitemOpen
	\bibfield  {author} {\bibinfo {author} {\bibfnamefont {M.}~\bibnamefont
			{Jamali}}, \bibinfo {author} {\bibfnamefont {J.~S.}\ \bibnamefont {Lee}},
		\bibinfo {author} {\bibfnamefont {J.~S.}\ \bibnamefont {Jeong}}, \bibinfo
		{author} {\bibfnamefont {F.}~\bibnamefont {Mahfouzi}}, \bibinfo {author}
		{\bibfnamefont {Y.}~\bibnamefont {Lv}}, \bibinfo {author} {\bibfnamefont
			{Z.}~\bibnamefont {Zhao}}, \bibinfo {author} {\bibfnamefont {B.~K.}\
			\bibnamefont {Nikoli\'c}}, \bibinfo {author} {\bibfnamefont {K.~A.}\
			\bibnamefont {Mkhoyan}}, \bibinfo {author} {\bibfnamefont {N.}~\bibnamefont
			{Samarth}}, \ and\ \bibinfo {author} {\bibfnamefont {J.-P.}\ \bibnamefont
			{Wang}},\ }\href {\doibase 10.1021/acs.nanolett.5b03274} {\bibfield
		{journal} {\bibinfo  {journal} {Nano Lett}\ }\textbf {\bibinfo {volume}
			{15}},\ \bibinfo {pages} {7126} (\bibinfo {year} {2015})}\BibitemShut
	{NoStop}%
	\bibitem [{\citenamefont {Kondou}\ \emph {et~al.}(2016)\citenamefont {Kondou},
		\citenamefont {Yoshimi}, \citenamefont {Tsukazaki}, \citenamefont {Fukuma},
		\citenamefont {Matsuno}, \citenamefont {Takahashi}, \citenamefont {Kawasaki},
		\citenamefont {Tokura},\ and\ \citenamefont {Otani}}]{Kondou2016}%
	\BibitemOpen
	\bibfield  {author} {\bibinfo {author} {\bibfnamefont {K.}~\bibnamefont
			{Kondou}}, \bibinfo {author} {\bibfnamefont {R.}~\bibnamefont {Yoshimi}},
		\bibinfo {author} {\bibfnamefont {A.}~\bibnamefont {Tsukazaki}}, \bibinfo
		{author} {\bibfnamefont {Y.}~\bibnamefont {Fukuma}}, \bibinfo {author}
		{\bibfnamefont {J.}~\bibnamefont {Matsuno}}, \bibinfo {author} {\bibfnamefont
			{K.~S.}\ \bibnamefont {Takahashi}}, \bibinfo {author} {\bibfnamefont
			{M.}~\bibnamefont {Kawasaki}}, \bibinfo {author} {\bibfnamefont
			{Y.}~\bibnamefont {Tokura}}, \ and\ \bibinfo {author} {\bibfnamefont
			{Y.}~\bibnamefont {Otani}},\ }\href {\doibase 10.1038/NPHYS3833} {\bibfield
		{journal} {\bibinfo  {journal} {Nature Physics}\ }\textbf {\bibinfo {volume}
			{12}},\ \bibinfo {pages} {1027} (\bibinfo {year} {2016})}\BibitemShut
	{NoStop}%
	\bibitem [{\citenamefont {Fanchiang}\ \emph {et~al.}(2018)\citenamefont
		{Fanchiang}, \citenamefont {Chen}, \citenamefont {Tseng}, \citenamefont
		{Chen}, \citenamefont {Cheng}, \citenamefont {Yang}, \citenamefont {Wu},
		\citenamefont {Lee}, \citenamefont {Hong},\ and\ \citenamefont
		{Kwo}}]{Fanchiang2018}%
	\BibitemOpen
	\bibfield  {author} {\bibinfo {author} {\bibfnamefont {Y.~T.}\ \bibnamefont
			{Fanchiang}}, \bibinfo {author} {\bibfnamefont {K.~H.~M.}\ \bibnamefont
			{Chen}}, \bibinfo {author} {\bibfnamefont {C.~C.}\ \bibnamefont {Tseng}},
		\bibinfo {author} {\bibfnamefont {C.~C.}\ \bibnamefont {Chen}}, \bibinfo
		{author} {\bibfnamefont {C.~K.}\ \bibnamefont {Cheng}}, \bibinfo {author}
		{\bibfnamefont {S.~R.}\ \bibnamefont {Yang}}, \bibinfo {author}
		{\bibfnamefont {C.~N.}\ \bibnamefont {Wu}}, \bibinfo {author} {\bibfnamefont
			{S.~F.}\ \bibnamefont {Lee}}, \bibinfo {author} {\bibfnamefont
			{M.}~\bibnamefont {Hong}}, \ and\ \bibinfo {author} {\bibfnamefont
			{J.}~\bibnamefont {Kwo}},\ }\href {\doibase 10.1038/s41467-017-02743-2}
	{\bibfield  {journal} {\bibinfo  {journal} {Nature Communications}\ }\textbf
		{\bibinfo {volume} {9}},\ \bibinfo {pages} {223} (\bibinfo {year}
		{2018})}\BibitemShut {NoStop}%
	\bibitem [{\citenamefont {Zhu}\ \emph {et~al.}(2021)\citenamefont {Zhu},
		\citenamefont {Wang}, \citenamefont {Shi}, \citenamefont {Teo}, \citenamefont
		{Wu},\ and\ \citenamefont {Yang}}]{zhu2021}%
	\BibitemOpen
	\bibfield  {author} {\bibinfo {author} {\bibfnamefont {D.}~\bibnamefont
			{Zhu}}, \bibinfo {author} {\bibfnamefont {Y.}~\bibnamefont {Wang}}, \bibinfo
		{author} {\bibfnamefont {S.}~\bibnamefont {Shi}}, \bibinfo {author}
		{\bibfnamefont {K.-L.}\ \bibnamefont {Teo}}, \bibinfo {author} {\bibfnamefont
			{Y.}~\bibnamefont {Wu}}, \ and\ \bibinfo {author} {\bibfnamefont
			{H.}~\bibnamefont {Yang}},\ }\href@noop {} {\bibfield  {journal} {\bibinfo
			{journal} {Applied Physics Letters}\ }\textbf {\bibinfo {volume} {118}},\
		\bibinfo {pages} {062403} (\bibinfo {year} {2021})}\BibitemShut {NoStop}%
	\bibitem [{\citenamefont {Ramaswamy}\ \emph {et~al.}(2019)\citenamefont
		{Ramaswamy}, \citenamefont {Dutta}, \citenamefont {Liang}, \citenamefont
		{Yang}, \citenamefont {Saifullah},\ and\ \citenamefont
		{Yang}}]{Ramaswamy2019}%
	\BibitemOpen
	\bibfield  {author} {\bibinfo {author} {\bibfnamefont {R.}~\bibnamefont
			{Ramaswamy}}, \bibinfo {author} {\bibfnamefont {T.}~\bibnamefont {Dutta}},
		\bibinfo {author} {\bibfnamefont {S.}~\bibnamefont {Liang}}, \bibinfo
		{author} {\bibfnamefont {G.}~\bibnamefont {Yang}}, \bibinfo {author}
		{\bibfnamefont {M.~S.~M.}\ \bibnamefont {Saifullah}}, \ and\ \bibinfo
		{author} {\bibfnamefont {H.}~\bibnamefont {Yang}},\ }\href {\doibase
		10.1088/1361-6463/ab0b96} {\bibfield  {journal} {\bibinfo  {journal} {Journal
				of Physics D: Applied Physics}\ }\textbf {\bibinfo {volume} {52}},\ \bibinfo
		{pages} {224001} (\bibinfo {year} {2019})}\BibitemShut {NoStop}%
	\bibitem [{\citenamefont {Baker}\ \emph {et~al.}(2015)\citenamefont {Baker},
		\citenamefont {Figueroa}, \citenamefont {Collins-McIntyre}, \citenamefont
		{Van Der~Laan},\ and\ \citenamefont {Hesjedal}}]{Baker2015}%
	\BibitemOpen
	\bibfield  {author} {\bibinfo {author} {\bibfnamefont {A.}~\bibnamefont
			{Baker}}, \bibinfo {author} {\bibfnamefont {A.}~\bibnamefont {Figueroa}},
		\bibinfo {author} {\bibfnamefont {L.}~\bibnamefont {Collins-McIntyre}},
		\bibinfo {author} {\bibfnamefont {G.}~\bibnamefont {Van Der~Laan}}, \ and\
		\bibinfo {author} {\bibfnamefont {T.}~\bibnamefont {Hesjedal}},\ }\href@noop
	{} {\bibfield  {journal} {\bibinfo  {journal} {Scientific reports}\ }\textbf
		{\bibinfo {volume} {5}},\ \bibinfo {pages} {1} (\bibinfo {year}
		{2015})}\BibitemShut {NoStop}%
	\bibitem [{\citenamefont {Deorani}\ \emph {et~al.}(2014)\citenamefont
		{Deorani}, \citenamefont {Son}, \citenamefont {Banerjee}, \citenamefont
		{Koirala}, \citenamefont {Brahlek}, \citenamefont {Oh},\ and\ \citenamefont
		{Yang}}]{Deorani2014}%
	\BibitemOpen
	\bibfield  {author} {\bibinfo {author} {\bibfnamefont {P.}~\bibnamefont
			{Deorani}}, \bibinfo {author} {\bibfnamefont {J.}~\bibnamefont {Son}},
		\bibinfo {author} {\bibfnamefont {K.}~\bibnamefont {Banerjee}}, \bibinfo
		{author} {\bibfnamefont {N.}~\bibnamefont {Koirala}}, \bibinfo {author}
		{\bibfnamefont {M.}~\bibnamefont {Brahlek}}, \bibinfo {author} {\bibfnamefont
			{S.}~\bibnamefont {Oh}}, \ and\ \bibinfo {author} {\bibfnamefont
			{H.}~\bibnamefont {Yang}},\ }\href {\doibase 10.1103/physrevb.90.094403}
	{\bibfield  {journal} {\bibinfo  {journal} {Physical Review B}\ }\textbf
		{\bibinfo {volume} {90}},\ \bibinfo{pages} {094403} (\bibinfo {year} {2014}),\
		10.1103/physrevb.90.094403}\BibitemShut {NoStop}%
	\bibitem [{\citenamefont {Shiomi}\ \emph {et~al.}(2014)\citenamefont {Shiomi},
		\citenamefont {Nomura}, \citenamefont {Kajiwara}, \citenamefont {Eto},
		\citenamefont {Novak}, \citenamefont {Segawa}, \citenamefont {Ando},\ and\
		\citenamefont {Saitoh}}]{Shiomi2014}%
	\BibitemOpen
	\bibfield  {author} {\bibinfo {author} {\bibfnamefont {Y.}~\bibnamefont
			{Shiomi}}, \bibinfo {author} {\bibfnamefont {K.}~\bibnamefont {Nomura}},
		\bibinfo {author} {\bibfnamefont {Y.}~\bibnamefont {Kajiwara}}, \bibinfo
		{author} {\bibfnamefont {K.}~\bibnamefont {Eto}}, \bibinfo {author}
		{\bibfnamefont {M.}~\bibnamefont {Novak}}, \bibinfo {author} {\bibfnamefont
			{K.}~\bibnamefont {Segawa}}, \bibinfo {author} {\bibfnamefont
			{Y.}~\bibnamefont {Ando}}, \ and\ \bibinfo {author} {\bibfnamefont
			{E.}~\bibnamefont {Saitoh}},\ }\href {\doibase
		10.1103/PhysRevLett.113.196601} {\bibfield  {journal} {\bibinfo  {journal}
			{Phys. Rev. Lett.}\ }\textbf {\bibinfo {volume} {113}},\ \bibinfo {pages}
		{196601} (\bibinfo {year} {2014})}\BibitemShut {NoStop}%
	\bibitem [{\citenamefont {Wang}\ \emph {et~al.}(2016)\citenamefont {Wang},
		\citenamefont {Kally}, \citenamefont {Lee}, \citenamefont {Liu},
		\citenamefont {Chang}, \citenamefont {Hickey}, \citenamefont {Mkhoyan},
		\citenamefont {Wu}, \citenamefont {Richardella},\ and\ \citenamefont
		{Samarth}}]{Wang2016}%
	\BibitemOpen
	\bibfield  {author} {\bibinfo {author} {\bibfnamefont {H.}~\bibnamefont
			{Wang}}, \bibinfo {author} {\bibfnamefont {J.}~\bibnamefont {Kally}},
		\bibinfo {author} {\bibfnamefont {J.~S.}\ \bibnamefont {Lee}}, \bibinfo
		{author} {\bibfnamefont {T.}~\bibnamefont {Liu}}, \bibinfo {author}
		{\bibfnamefont {H.}~\bibnamefont {Chang}}, \bibinfo {author} {\bibfnamefont
			{D.~R.}\ \bibnamefont {Hickey}}, \bibinfo {author} {\bibfnamefont {K.~A.}\
			\bibnamefont {Mkhoyan}}, \bibinfo {author} {\bibfnamefont {M.}~\bibnamefont
			{Wu}}, \bibinfo {author} {\bibfnamefont {A.}~\bibnamefont {Richardella}}, \
		and\ \bibinfo {author} {\bibfnamefont {N.}~\bibnamefont {Samarth}},\ }\href
	{\doibase 10.1103/PhysRevLett.117.076601} {\bibfield  {journal} {\bibinfo
			{journal} {Phys. Rev. Lett.}\ }\textbf {\bibinfo {volume} {117}},\ \bibinfo
		{pages} {076601} (\bibinfo {year} {2016})}\BibitemShut {NoStop}%
	\bibitem [{\citenamefont {Singh}\ \emph {et~al.}(2020)\citenamefont {Singh},
		\citenamefont {Jena}, \citenamefont {Samanta}, \citenamefont {Biswas},\ and\
		\citenamefont {Bedanta}}]{singh2020}%
	\BibitemOpen
	\bibfield  {author} {\bibinfo {author} {\bibfnamefont {B.~B.}\ \bibnamefont
			{Singh}}, \bibinfo {author} {\bibfnamefont {S.~K.}\ \bibnamefont {Jena}},
		\bibinfo {author} {\bibfnamefont {M.}~\bibnamefont {Samanta}}, \bibinfo
		{author} {\bibfnamefont {K.}~\bibnamefont {Biswas}}, \ and\ \bibinfo {author}
		{\bibfnamefont {S.}~\bibnamefont {Bedanta}},\ }\href@noop {} {\bibfield
		{journal} {\bibinfo  {journal} {ACS Applied Materials \& Interfaces}\
		}\textbf {\bibinfo {volume} {12}},\ \bibinfo {pages} {53409} (\bibinfo {year}
		{2020})}\BibitemShut {NoStop}%
	\bibitem [{\citenamefont {Zheng}\ \emph {et~al.}(2020)\citenamefont {Zheng},
		\citenamefont {Zhang}, \citenamefont {Zhu}, \citenamefont {Zhang},
		\citenamefont {Feng}, \citenamefont {He}, \citenamefont {Chen}, \citenamefont
		{Zhang}, \citenamefont {Liu}, \citenamefont {Zhang} \emph
		{et~al.}}]{zheng2020}%
	\BibitemOpen
	\bibfield  {author} {\bibinfo {author} {\bibfnamefont {Z.}~\bibnamefont
			{Zheng}}, \bibinfo {author} {\bibfnamefont {Y.}~\bibnamefont {Zhang}},
		\bibinfo {author} {\bibfnamefont {D.}~\bibnamefont {Zhu}}, \bibinfo {author}
		{\bibfnamefont {K.}~\bibnamefont {Zhang}}, \bibinfo {author} {\bibfnamefont
			{X.}~\bibnamefont {Feng}}, \bibinfo {author} {\bibfnamefont {Y.}~\bibnamefont
			{He}}, \bibinfo {author} {\bibfnamefont {L.}~\bibnamefont {Chen}}, \bibinfo
		{author} {\bibfnamefont {Z.}~\bibnamefont {Zhang}}, \bibinfo {author}
		{\bibfnamefont {D.}~\bibnamefont {Liu}}, \bibinfo {author} {\bibfnamefont
			{Y.}~\bibnamefont {Zhang}},  \emph {et~al.},\ }\href@noop {} {\bibfield
		{journal} {\bibinfo  {journal} {Chinese Physics B}\ }\textbf {\bibinfo
			{volume} {29}},\ \bibinfo {pages} {078505} (\bibinfo {year}
		{2020})}\BibitemShut {NoStop}%
	\bibitem [{\citenamefont {Fan}\ \emph {et~al.}(2022)\citenamefont {Fan},
		\citenamefont {Khang}, \citenamefont {Nakano},\ and\ \citenamefont
		{Hai}}]{Fan2022}%
	\BibitemOpen
	\bibfield  {author} {\bibinfo {author} {\bibfnamefont {T.}~\bibnamefont
			{Fan}}, \bibinfo {author} {\bibfnamefont {N.~H.~D.}\ \bibnamefont {Khang}},
		\bibinfo {author} {\bibfnamefont {S.}~\bibnamefont {Nakano}}, \ and\ \bibinfo
		{author} {\bibfnamefont {P.~N.}\ \bibnamefont {Hai}},\ }\href@noop {}
	{\bibfield  {journal} {\bibinfo  {journal} {Scientific reports}\ }\textbf
		{\bibinfo {volume} {12}},\ \bibinfo {pages} {1} (\bibinfo {year}
		{2022})}\BibitemShut {NoStop}%
	\bibitem [{\citenamefont {Fan}\ \emph {et~al.}(2014)\citenamefont {Fan},
		\citenamefont {Upadhyaya}, \citenamefont {Kou}, \citenamefont {Lang},
		\citenamefont {Takei}, \citenamefont {Wang}, \citenamefont {Tang},
		\citenamefont {He}, \citenamefont {Chang}, \citenamefont {Montazeri},
		\citenamefont {Yu}, \citenamefont {Jiang}, \citenamefont {Nie}, \citenamefont
		{Schwartz}, \citenamefont {Tserkovnyak},\ and\ \citenamefont
		{Wang}}]{Fan2014}%
	\BibitemOpen
	\bibfield  {author} {\bibinfo {author} {\bibfnamefont {Y.}~\bibnamefont
			{Fan}}, \bibinfo {author} {\bibfnamefont {P.}~\bibnamefont {Upadhyaya}},
		\bibinfo {author} {\bibfnamefont {X.}~\bibnamefont {Kou}}, \bibinfo {author}
		{\bibfnamefont {M.}~\bibnamefont {Lang}}, \bibinfo {author} {\bibfnamefont
			{S.}~\bibnamefont {Takei}}, \bibinfo {author} {\bibfnamefont
			{Z.}~\bibnamefont {Wang}}, \bibinfo {author} {\bibfnamefont {J.}~\bibnamefont
			{Tang}}, \bibinfo {author} {\bibfnamefont {L.}~\bibnamefont {He}}, \bibinfo
		{author} {\bibfnamefont {L.-T.}\ \bibnamefont {Chang}}, \bibinfo {author}
		{\bibfnamefont {M.}~\bibnamefont {Montazeri}}, \bibinfo {author}
		{\bibfnamefont {G.}~\bibnamefont {Yu}}, \bibinfo {author} {\bibfnamefont
			{W.}~\bibnamefont {Jiang}}, \bibinfo {author} {\bibfnamefont
			{T.}~\bibnamefont {Nie}}, \bibinfo {author} {\bibfnamefont {R.~N.}\
			\bibnamefont {Schwartz}}, \bibinfo {author} {\bibfnamefont {Y.}~\bibnamefont
			{Tserkovnyak}}, \ and\ \bibinfo {author} {\bibfnamefont {K.~L.}\ \bibnamefont
			{Wang}},\ }\href {\doibase 10.1038/NMAT3973} {\bibfield  {journal} {\bibinfo
			{journal} {Nature Materials}\ }\textbf {\bibinfo {volume} {13}},\ \bibinfo
		{pages} {699} (\bibinfo {year} {2014})}\BibitemShut {NoStop}%
	\bibitem [{\citenamefont {Fan}\ \emph {et~al.}(2016)\citenamefont {Fan},
		\citenamefont {Kou}, \citenamefont {Upadhyaya}, \citenamefont {Shao},
		\citenamefont {Pan}, \citenamefont {Lang}, \citenamefont {Che}, \citenamefont
		{Tang}, \citenamefont {Montazeri}, \citenamefont {Murata} \emph
		{et~al.}}]{fan2016}%
	\BibitemOpen
	\bibfield  {author} {\bibinfo {author} {\bibfnamefont {Y.}~\bibnamefont
			{Fan}}, \bibinfo {author} {\bibfnamefont {X.}~\bibnamefont {Kou}}, \bibinfo
		{author} {\bibfnamefont {P.}~\bibnamefont {Upadhyaya}}, \bibinfo {author}
		{\bibfnamefont {Q.}~\bibnamefont {Shao}}, \bibinfo {author} {\bibfnamefont
			{L.}~\bibnamefont {Pan}}, \bibinfo {author} {\bibfnamefont {M.}~\bibnamefont
			{Lang}}, \bibinfo {author} {\bibfnamefont {X.}~\bibnamefont {Che}}, \bibinfo
		{author} {\bibfnamefont {J.}~\bibnamefont {Tang}}, \bibinfo {author}
		{\bibfnamefont {M.}~\bibnamefont {Montazeri}}, \bibinfo {author}
		{\bibfnamefont {K.}~\bibnamefont {Murata}},  \emph {et~al.},\ }\href@noop {}
	{\bibfield  {journal} {\bibinfo  {journal} {Nature nanotechnology}\ }\textbf
		{\bibinfo {volume} {11}},\ \bibinfo {pages} {352} (\bibinfo {year}
		{2016})}\BibitemShut {NoStop}%
	\bibitem [{\citenamefont {Chang}\ \emph {et~al.}(2015)\citenamefont {Chang},
		\citenamefont {Markussen}, \citenamefont {Smidstrup}, \citenamefont
		{Stokbro},\ and\ \citenamefont {Nikoli{\'{c}}}}]{Chang2015}%
	\BibitemOpen
	\bibfield  {author} {\bibinfo {author} {\bibfnamefont {P.~H.}\ \bibnamefont
			{Chang}}, \bibinfo {author} {\bibfnamefont {T.}~\bibnamefont {Markussen}},
		\bibinfo {author} {\bibfnamefont {S.}~\bibnamefont {Smidstrup}}, \bibinfo
		{author} {\bibfnamefont {K.}~\bibnamefont {Stokbro}}, \ and\ \bibinfo
		{author} {\bibfnamefont {B.~K.}\ \bibnamefont {Nikoli{\'{c}}}},\ }\href
	{\doibase 10.1103/PHYSREVB.92.201406/FIGURES/4/MEDIUM} {\bibfield  {journal}
		{\bibinfo  {journal} {Physical Review B - Condensed Matter and Materials
				Physics}\ }\textbf {\bibinfo {volume} {92}},\ \bibinfo {pages} {201406(R)}
		(\bibinfo {year} {2015})},\ \Eprint {http://arxiv.org/abs/1503.08046}
	{arXiv:1503.08046} \BibitemShut {NoStop}%
	\bibitem [{\citenamefont {Ghosh}\ and\ \citenamefont
		{Manchon}(2018)}]{Ghosh2018}%
	\BibitemOpen
	\bibfield  {author} {\bibinfo {author} {\bibfnamefont {S.}~\bibnamefont
			{Ghosh}}\ and\ \bibinfo {author} {\bibfnamefont {A.}~\bibnamefont
			{Manchon}},\ }\href {\doibase 10.1103/PHYSREVB.97.134402/FIGURES/10/MEDIUM}
	{\bibfield  {journal} {\bibinfo  {journal} {Physical Review B}\ }\textbf
		{\bibinfo {volume} {97}},\ \bibinfo {pages} {134402} (\bibinfo {year}
		{2018})}\BibitemShut {NoStop}%
	\bibitem [{\citenamefont {Ado}\ \emph {et~al.}(2017)\citenamefont {Ado},
		\citenamefont {Tretiakov},\ and\ \citenamefont {Titov}}]{Ado2017}%
	\BibitemOpen
	\bibfield  {author} {\bibinfo {author} {\bibfnamefont {I.~A.}~\bibnamefont
			{Ado}}, \bibinfo {author} {\bibfnamefont {O.~A.}\ \bibnamefont {Tretiakov}},
		\ and\ \bibinfo {author} {\bibfnamefont {M.}~\bibnamefont {Titov}},\
	}\href@noop {} {\bibfield  {journal} {\bibinfo  {journal} {Physical Review
				B}\ }\textbf {\bibinfo {volume} {95}},\ \bibinfo {pages} {094401} (\bibinfo
		{year} {2017})}\BibitemShut {NoStop}%
	\bibitem [{\citenamefont {Kurebayashi}\ and\ \citenamefont
		{Nagaosa}(2019)}]{Kurebayashi2019}%
	\BibitemOpen
	\bibfield  {author} {\bibinfo {author} {\bibfnamefont {D.}~\bibnamefont
			{Kurebayashi}}\ and\ \bibinfo {author} {\bibfnamefont {N.}~\bibnamefont
			{Nagaosa}},\ }\href {\doibase 10.1103/PhysRevB.100.134407} {\bibfield
		{journal} {\bibinfo  {journal} {Physical Review B}\ }\textbf {\bibinfo
			{volume} {100}},\ \bibinfo {pages} {134407} (\bibinfo {year}
		{2019})}\BibitemShut {NoStop}%
	\bibitem [{\citenamefont {Gao}\ \emph {et~al.}(2019)\citenamefont {Gao},
		\citenamefont {Tazaki}, \citenamefont {Asami}, \citenamefont {Nakayama},\
		and\ \citenamefont {Ando}}]{gao2019}%
	\BibitemOpen
	\bibfield  {author} {\bibinfo {author} {\bibfnamefont {T.}~\bibnamefont
			{Gao}}, \bibinfo {author} {\bibfnamefont {Y.}~\bibnamefont {Tazaki}},
		\bibinfo {author} {\bibfnamefont {A.}~\bibnamefont {Asami}}, \bibinfo
		{author} {\bibfnamefont {H.}~\bibnamefont {Nakayama}}, \ and\ \bibinfo
		{author} {\bibfnamefont {K.}~\bibnamefont {Ando}},\ }\href@noop {} {\bibfield
		{journal} {\bibinfo  {journal} {arXiv preprint arXiv:1911.00413}\ }
		(\bibinfo {year} {2019})}\BibitemShut {NoStop}%
	\bibitem [{\citenamefont {Wang}\ \emph {et~al.}(2018)\citenamefont {Wang},
		\citenamefont {Ramaswamy},\ and\ \citenamefont {Yang}}]{Wang2018}%
	\BibitemOpen
	\bibfield  {author} {\bibinfo {author} {\bibfnamefont {Y.}~\bibnamefont
			{Wang}}, \bibinfo {author} {\bibfnamefont {R.}~\bibnamefont {Ramaswamy}}, \
		and\ \bibinfo {author} {\bibfnamefont {H.}~\bibnamefont {Yang}},\ }\href
	{\doibase 10.1088/1361-6463/aac7b5} {\bibfield  {journal} {\bibinfo
			{journal} {Journal of Physics D: Applied Physics}\ }\textbf {\bibinfo
			{volume} {51}},\ \bibinfo {pages} {273002} (\bibinfo {year}
		{2018})}\BibitemShut {NoStop}%
	\bibitem [{\citenamefont {Zhang}\ \emph {et~al.}(2016)\citenamefont {Zhang},
		\citenamefont {Velev}, \citenamefont {Dang},\ and\ \citenamefont
		{Tsymbal}}]{Zhang2016}%
	\BibitemOpen
	\bibfield  {author} {\bibinfo {author} {\bibfnamefont {J.}~\bibnamefont
			{Zhang}}, \bibinfo {author} {\bibfnamefont {J.~P.}\ \bibnamefont {Velev}},
		\bibinfo {author} {\bibfnamefont {X.}~\bibnamefont {Dang}}, \ and\ \bibinfo
		{author} {\bibfnamefont {E.~Y.}\ \bibnamefont {Tsymbal}},\ }\href {\doibase
		10.1103/PhysRevB.94.014435} {\bibfield  {journal} {\bibinfo  {journal}
			{Physical Review B}\ }\textbf {\bibinfo {volume} {94}},\ \bibinfo {pages}
		{014435} (\bibinfo {year} {2016})}\BibitemShut {NoStop}%
	\bibitem [{\citenamefont {Marmolejo-Tejada}\ \emph {et~al.}(2017)\citenamefont
		{Marmolejo-Tejada}, \citenamefont {Dolui}, \citenamefont {Lazi{\'{c}}},
		\citenamefont {Chang}, \citenamefont {Smidstrup}, \citenamefont {Stradi},
		\citenamefont {Stokbro},\ and\ \citenamefont
		{Nikoli{\'{c}}}}]{Marmolejo-Tejada2017}%
	\BibitemOpen
	\bibfield  {author} {\bibinfo {author} {\bibfnamefont {J.~M.}\ \bibnamefont
			{Marmolejo-Tejada}}, \bibinfo {author} {\bibfnamefont {K.}~\bibnamefont
			{Dolui}}, \bibinfo {author} {\bibfnamefont {P.}~\bibnamefont {Lazi{\'{c}}}},
		\bibinfo {author} {\bibfnamefont {P.~H.}\ \bibnamefont {Chang}}, \bibinfo
		{author} {\bibfnamefont {S.}~\bibnamefont {Smidstrup}}, \bibinfo {author}
		{\bibfnamefont {D.}~\bibnamefont {Stradi}}, \bibinfo {author} {\bibfnamefont
			{K.}~\bibnamefont {Stokbro}}, \ and\ \bibinfo {author} {\bibfnamefont
			{B.~K.}\ \bibnamefont {Nikoli{\'{c}}}},\ }\href {\doibase
		10.1021/ACS.NANOLETT.7B02511} {\bibfield  {journal} {\bibinfo  {journal}
			{Nano Letters}\ }\textbf {\bibinfo {volume} {17}},\ \bibinfo {pages} {5626}
		(\bibinfo {year} {2017})},\ \Eprint {http://arxiv.org/abs/1701.00462}
	{arXiv:1701.00462} \BibitemShut {NoStop}%
	\bibitem [{\citenamefont {Jash}\ \emph {et~al.}(2021)\citenamefont {Jash},
		\citenamefont {Kumar}, \citenamefont {Ghosh}, \citenamefont {Bharathi},\ and\
		\citenamefont {Banerjee}}]{Jash2021}%
	\BibitemOpen
	\bibfield  {author} {\bibinfo {author} {\bibfnamefont {A.}~\bibnamefont
			{Jash}}, \bibinfo {author} {\bibfnamefont {A.}~\bibnamefont {Kumar}},
		\bibinfo {author} {\bibfnamefont {S.}~\bibnamefont {Ghosh}}, \bibinfo
		{author} {\bibfnamefont {A.}~\bibnamefont {Bharathi}}, \ and\ \bibinfo
		{author} {\bibfnamefont {S.~S.}\ \bibnamefont {Banerjee}},\ }\href {\doibase
		10.1038/s41598-021-86706-0} {\bibfield  {journal} {\bibinfo  {journal}
			{Scientific Reports}\ }\textbf {\bibinfo {volume} {11}},\ \bibinfo {pages}
		{7445} (\bibinfo {year} {2021})}\BibitemShut {NoStop}%
	\bibitem [{\citenamefont {Siu}\ \emph {et~al.}(2018)\citenamefont {Siu},
		\citenamefont {Wang}, \citenamefont {Yang},\ and\ \citenamefont
		{Jalil}}]{siu2018}%
	\BibitemOpen
	\bibfield  {author} {\bibinfo {author} {\bibfnamefont {Z.~B.}\ \bibnamefont
			{Siu}}, \bibinfo {author} {\bibfnamefont {Y.}~\bibnamefont {Wang}}, \bibinfo
		{author} {\bibfnamefont {H.}~\bibnamefont {Yang}}, \ and\ \bibinfo {author}
		{\bibfnamefont {M.~B.}\ \bibnamefont {Jalil}},\ }\href@noop {} {\bibfield
		{journal} {\bibinfo  {journal} {Journal of Physics D: Applied Physics}\
		}\textbf {\bibinfo {volume} {51}},\ \bibinfo {pages} {425301} (\bibinfo
		{year} {2018})}\BibitemShut {NoStop}%
	\bibitem [{\citenamefont {Vobornik}\ \emph {et~al.}(2011)\citenamefont
		{Vobornik}, \citenamefont {Manju}, \citenamefont {Fujii}, \citenamefont
		{Borgatti}, \citenamefont {Torelli}, \citenamefont {Krizmancic},
		\citenamefont {Hor}, \citenamefont {Cava},\ and\ \citenamefont
		{Panaccione}}]{Vobornik2011}%
	\BibitemOpen
	\bibfield  {author} {\bibinfo {author} {\bibfnamefont {I.}~\bibnamefont
			{Vobornik}}, \bibinfo {author} {\bibfnamefont {U.}~\bibnamefont {Manju}},
		\bibinfo {author} {\bibfnamefont {J.}~\bibnamefont {Fujii}}, \bibinfo
		{author} {\bibfnamefont {F.}~\bibnamefont {Borgatti}}, \bibinfo {author}
		{\bibfnamefont {P.}~\bibnamefont {Torelli}}, \bibinfo {author} {\bibfnamefont
			{D.}~\bibnamefont {Krizmancic}}, \bibinfo {author} {\bibfnamefont {Y.~S.}\
			\bibnamefont {Hor}}, \bibinfo {author} {\bibfnamefont {R.~J.}\ \bibnamefont
			{Cava}}, \ and\ \bibinfo {author} {\bibfnamefont {G.}~\bibnamefont
			{Panaccione}},\ }\href {\doibase 10.1021/nl201275q} {\bibfield  {journal}
		{\bibinfo  {journal} {Nano Letters}\ }\textbf {\bibinfo {volume} {11}},\
		\bibinfo {pages} {4079} (\bibinfo {year} {2011})}\BibitemShut {NoStop}%
	\bibitem [{\citenamefont {Eremeev}\ \emph {et~al.}(2013)\citenamefont
		{Eremeev}, \citenamefont {Men'shov}, \citenamefont {Tugushev}, \citenamefont
		{Echenique},\ and\ \citenamefont {Chulkov}}]{Eremeev2013}%
	\BibitemOpen
	\bibfield  {author} {\bibinfo {author} {\bibfnamefont {S.~V.}\ \bibnamefont
			{Eremeev}}, \bibinfo {author} {\bibfnamefont {V.~N.}\ \bibnamefont
			{Men'shov}}, \bibinfo {author} {\bibfnamefont {V.~V.}\ \bibnamefont
			{Tugushev}}, \bibinfo {author} {\bibfnamefont {P.~M.}\ \bibnamefont
			{Echenique}}, \ and\ \bibinfo {author} {\bibfnamefont {E.~V.}\ \bibnamefont
			{Chulkov}},\ }\href {\doibase 10.1103/PhysRevB.88.144430} {\bibfield
		{journal} {\bibinfo  {journal} {Phys. Rev. B}\ }\textbf {\bibinfo {volume}
			{88}},\ \bibinfo {pages} {144430} (\bibinfo {year} {2013})}\BibitemShut
	{NoStop}%
	\bibitem [{\citenamefont {Lang}\ \emph {et~al.}(2014)\citenamefont {Lang},
		\citenamefont {Montazeri}, \citenamefont {Onbasli}, \citenamefont {Kou},
		\citenamefont {Fan}, \citenamefont {Upadhyaya}, \citenamefont {Yao},
		\citenamefont {Liu}, \citenamefont {Jiang}, \citenamefont {Jiang},
		\citenamefont {Wong}, \citenamefont {Yu}, \citenamefont {Tang}, \citenamefont
		{Nie}, \citenamefont {He}, \citenamefont {Schwartz}, \citenamefont {Wang},
		\citenamefont {Ross},\ and\ \citenamefont {Wang}}]{Lang2014}%
	\BibitemOpen
	\bibfield  {author} {\bibinfo {author} {\bibfnamefont {M.}~\bibnamefont
			{Lang}}, \bibinfo {author} {\bibfnamefont {M.}~\bibnamefont {Montazeri}},
		\bibinfo {author} {\bibfnamefont {M.~C.}\ \bibnamefont {Onbasli}}, \bibinfo
		{author} {\bibfnamefont {X.}~\bibnamefont {Kou}}, \bibinfo {author}
		{\bibfnamefont {Y.}~\bibnamefont {Fan}}, \bibinfo {author} {\bibfnamefont
			{P.}~\bibnamefont {Upadhyaya}}, \bibinfo {author} {\bibfnamefont
			{K.}~\bibnamefont {Yao}}, \bibinfo {author} {\bibfnamefont {F.}~\bibnamefont
			{Liu}}, \bibinfo {author} {\bibfnamefont {Y.}~\bibnamefont {Jiang}}, \bibinfo
		{author} {\bibfnamefont {W.}~\bibnamefont {Jiang}}, \bibinfo {author}
		{\bibfnamefont {K.~L.}\ \bibnamefont {Wong}}, \bibinfo {author}
		{\bibfnamefont {G.}~\bibnamefont {Yu}}, \bibinfo {author} {\bibfnamefont
			{J.}~\bibnamefont {Tang}}, \bibinfo {author} {\bibfnamefont {T.}~\bibnamefont
			{Nie}}, \bibinfo {author} {\bibfnamefont {L.}~\bibnamefont {He}}, \bibinfo
		{author} {\bibfnamefont {R.~N.}\ \bibnamefont {Schwartz}}, \bibinfo {author}
		{\bibfnamefont {Y.}~\bibnamefont {Wang}}, \bibinfo {author} {\bibfnamefont
			{C.~A.}\ \bibnamefont {Ross}}, \ and\ \bibinfo {author} {\bibfnamefont
			{K.~L.}\ \bibnamefont {Wang}},\ }\href {\doibase 10.1021/nl500973k}
	{\bibfield  {journal} {\bibinfo  {journal} {Nano Letters}\ }\textbf {\bibinfo
			{volume} {14}},\ \bibinfo {pages} {3459} (\bibinfo {year} {2014})},\ \bibinfo
	{note} {pMID: 24844837},\ \Eprint
	{http://arxiv.org/abs/https://doi.org/10.1021/nl500973k}
	{https://doi.org/10.1021/nl500973k} \BibitemShut {NoStop}%
	\bibitem [{\citenamefont {Katmis}\ \emph {et~al.}(2016)\citenamefont {Katmis},
		\citenamefont {Lauter}, \citenamefont {Nogueira}, \citenamefont {Assaf},
		\citenamefont {Jamer}, \citenamefont {Wei}, \citenamefont {Satpati},
		\citenamefont {Freeland}, \citenamefont {Eremin}, \citenamefont {Heiman}
		\emph {et~al.}}]{katmis2016}%
	\BibitemOpen
	\bibfield  {author} {\bibinfo {author} {\bibfnamefont {F.}~\bibnamefont
			{Katmis}}, \bibinfo {author} {\bibfnamefont {V.}~\bibnamefont {Lauter}},
		\bibinfo {author} {\bibfnamefont {F.~S.}\ \bibnamefont {Nogueira}}, \bibinfo
		{author} {\bibfnamefont {B.~A.}\ \bibnamefont {Assaf}}, \bibinfo {author}
		{\bibfnamefont {M.~E.}\ \bibnamefont {Jamer}}, \bibinfo {author}
		{\bibfnamefont {P.}~\bibnamefont {Wei}}, \bibinfo {author} {\bibfnamefont
			{B.}~\bibnamefont {Satpati}}, \bibinfo {author} {\bibfnamefont {J.~W.}\
			\bibnamefont {Freeland}}, \bibinfo {author} {\bibfnamefont {I.}~\bibnamefont
			{Eremin}}, \bibinfo {author} {\bibfnamefont {D.}~\bibnamefont {Heiman}},
		\emph {et~al.},\ }\href@noop {} {\bibfield  {journal} {\bibinfo  {journal}
			{Nature}\ }\textbf {\bibinfo {volume} {533}},\ \bibinfo {pages} {513}
		(\bibinfo {year} {2016})}\BibitemShut {NoStop}%
	\bibitem [{\citenamefont {Lee}\ \emph {et~al.}(2016)\citenamefont {Lee},
		\citenamefont {Katmis}, \citenamefont {Jarillo-Herrero}, \citenamefont
		{Moodera},\ and\ \citenamefont {Gedik}}]{lee2016}%
	\BibitemOpen
	\bibfield  {author} {\bibinfo {author} {\bibfnamefont {C.}~\bibnamefont
			{Lee}}, \bibinfo {author} {\bibfnamefont {F.}~\bibnamefont {Katmis}},
		\bibinfo {author} {\bibfnamefont {P.}~\bibnamefont {Jarillo-Herrero}},
		\bibinfo {author} {\bibfnamefont {J.~S.}\ \bibnamefont {Moodera}}, \ and\
		\bibinfo {author} {\bibfnamefont {N.}~\bibnamefont {Gedik}},\ }\href@noop {}
	{\bibfield  {journal} {\bibinfo  {journal} {Nature communications}\ }\textbf
		{\bibinfo {volume} {7}},\ \bibinfo {pages} {1} (\bibinfo {year}
		{2016})}\BibitemShut {NoStop}%
	\bibitem [{\citenamefont {Wei}\ \emph {et~al.}(2013)\citenamefont {Wei},
		\citenamefont {Katmis}, \citenamefont {Assaf}, \citenamefont {Steinberg},
		\citenamefont {Jarillo-Herrero}, \citenamefont {Heiman},\ and\ \citenamefont
		{Moodera}}]{wei2013}%
	\BibitemOpen
	\bibfield  {author} {\bibinfo {author} {\bibfnamefont {P.}~\bibnamefont
			{Wei}}, \bibinfo {author} {\bibfnamefont {F.}~\bibnamefont {Katmis}},
		\bibinfo {author} {\bibfnamefont {B.~A.}\ \bibnamefont {Assaf}}, \bibinfo
		{author} {\bibfnamefont {H.}~\bibnamefont {Steinberg}}, \bibinfo {author}
		{\bibfnamefont {P.}~\bibnamefont {Jarillo-Herrero}}, \bibinfo {author}
		{\bibfnamefont {D.}~\bibnamefont {Heiman}}, \ and\ \bibinfo {author}
		{\bibfnamefont {J.~S.}\ \bibnamefont {Moodera}},\ }\href@noop {} {\bibfield
		{journal} {\bibinfo  {journal} {Physical review letters}\ }\textbf {\bibinfo
			{volume} {110}},\ \bibinfo {pages} {186807} (\bibinfo {year}
		{2013})}\BibitemShut {NoStop}%
	\bibitem [{\citenamefont {Kandala}\ \emph {et~al.}(2013)\citenamefont
		{Kandala}, \citenamefont {Richardella}, \citenamefont {Rench}, \citenamefont
		{Zhang}, \citenamefont {Flanagan},\ and\ \citenamefont
		{Samarth}}]{kandala2013}%
	\BibitemOpen
	\bibfield  {author} {\bibinfo {author} {\bibfnamefont {A.}~\bibnamefont
			{Kandala}}, \bibinfo {author} {\bibfnamefont {A.}~\bibnamefont
			{Richardella}}, \bibinfo {author} {\bibfnamefont {D.}~\bibnamefont {Rench}},
		\bibinfo {author} {\bibfnamefont {D.}~\bibnamefont {Zhang}}, \bibinfo
		{author} {\bibfnamefont {T.}~\bibnamefont {Flanagan}}, \ and\ \bibinfo
		{author} {\bibfnamefont {N.}~\bibnamefont {Samarth}},\ }\href@noop {}
	{\bibfield  {journal} {\bibinfo  {journal} {Applied Physics Letters}\
		}\textbf {\bibinfo {volume} {103}},\ \bibinfo {pages} {202409} (\bibinfo
		{year} {2013})}\BibitemShut {NoStop}%
	\bibitem [{\citenamefont {Culcer}\ and\ \citenamefont
		{Winkler}(2007)}]{Culcer2007}%
	\BibitemOpen
	\bibfield  {author} {\bibinfo {author} {\bibfnamefont {D.}~\bibnamefont
			{Culcer}}\ and\ \bibinfo {author} {\bibfnamefont {R.}~\bibnamefont
			{Winkler}},\ }\href {\doibase 10.1103/PhysRevB.76.245322} {\bibfield
		{journal} {\bibinfo  {journal} {Phys. Rev. B}\ }\textbf {\bibinfo {volume}
			{76}},\ \bibinfo {pages} {245322} (\bibinfo {year} {2007})}\BibitemShut
	{NoStop}%
	\bibitem [{\citenamefont {Bi}\ \emph {et~al.}(2013)\citenamefont {Bi},
		\citenamefont {He}, \citenamefont {Hankiewicz}, \citenamefont {Winkler},
		\citenamefont {Vignale},\ and\ \citenamefont {Culcer}}]{Bi2013}%
	\BibitemOpen
	\bibfield  {author} {\bibinfo {author} {\bibfnamefont {X.}~\bibnamefont
			{Bi}}, \bibinfo {author} {\bibfnamefont {P.}~\bibnamefont {He}}, \bibinfo
		{author} {\bibfnamefont {E.-M.}~\bibnamefont {Hankiewicz}}, \bibinfo {author}
		{\bibfnamefont {R.}~\bibnamefont {Winkler}}, \bibinfo {author} {\bibfnamefont
			{G.}~\bibnamefont {Vignale}}, \ and\ \bibinfo {author} {\bibfnamefont
			{D.}~\bibnamefont {Culcer}},\ }\href@noop {} {\bibfield  {journal} {\bibinfo
			{journal} {Physical Review B}\ }\textbf {\bibinfo {volume} {88}},\ \bibinfo
		{pages} {035316} (\bibinfo {year} {2013})}\BibitemShut {NoStop}%
	\bibitem [{\citenamefont {Culcer}\ \emph {et~al.}(2017)\citenamefont {Culcer},
		\citenamefont {Sekine},\ and\ \citenamefont {MacDonald}}]{Culcer2017}%
	\BibitemOpen
	\bibfield  {author} {\bibinfo {author} {\bibfnamefont {D.}~\bibnamefont
			{Culcer}}, \bibinfo {author} {\bibfnamefont {A.}~\bibnamefont {Sekine}}, \
		and\ \bibinfo {author} {\bibfnamefont {A.~H.}\ \bibnamefont {MacDonald}},\
	}\href@noop {} {\bibfield  {journal} {\bibinfo  {journal} {Physical Review
				B}\ }\textbf {\bibinfo {volume} {96}},\ \bibinfo {pages} {035106} (\bibinfo
		{year} {2017})}\BibitemShut {NoStop}%
	\bibitem [{\citenamefont {Culcer}\ \emph {et~al.}(2009)\citenamefont {Culcer},
		\citenamefont {Lucassen}, \citenamefont {Duine},\ and\ \citenamefont
		{Winkler}}]{culcer2009}%
	\BibitemOpen
	\bibfield  {author} {\bibinfo {author} {\bibfnamefont {D.}~\bibnamefont
			{Culcer}}, \bibinfo {author} {\bibfnamefont {M.~E.}~\bibnamefont {Lucassen}},
		\bibinfo {author} {\bibfnamefont {R.~A.}~\bibnamefont {Duine}}, \ and\ \bibinfo
		{author} {\bibfnamefont {R.}~\bibnamefont {Winkler}},\ }\href@noop {}
	{\bibfield  {journal} {\bibinfo  {journal} {Physical Review B}\ }\textbf
		{\bibinfo {volume} {79}},\ \bibinfo {pages} {155208} (\bibinfo {year}
		{2009})}\BibitemShut {NoStop}%
	\bibitem [{\citenamefont {Winkler}\ \emph {et~al.}(2008)\citenamefont
		{Winkler}, \citenamefont {Culcer}, \citenamefont {Papadakis}, \citenamefont
		{Habib},\ and\ \citenamefont {Shayegan}}]{winkler2008}%
	\BibitemOpen
	\bibfield  {author} {\bibinfo {author} {\bibfnamefont {R.}~\bibnamefont
			{Winkler}}, \bibinfo {author} {\bibfnamefont {D.}~\bibnamefont {Culcer}},
		\bibinfo {author} {\bibfnamefont {S.}~\bibnamefont {Papadakis}}, \bibinfo
		{author} {\bibfnamefont {B.}~\bibnamefont {Habib}}, \ and\ \bibinfo {author}
		{\bibfnamefont {M.}~\bibnamefont {Shayegan}},\ }\href@noop {} {\bibfield
		{journal} {\bibinfo  {journal} {Semiconductor science and technology}\
		}\textbf {\bibinfo {volume} {23}},\ \bibinfo {pages} {114017} (\bibinfo
		{year} {2008})}\BibitemShut {NoStop}%
	\bibitem [{\citenamefont {Sekine}\ and\ \citenamefont
		{MacDonald}(2018)}]{sekine2018}%
	\BibitemOpen
	\bibfield  {author} {\bibinfo {author} {\bibfnamefont {A.}~\bibnamefont
			{Sekine}}\ and\ \bibinfo {author} {\bibfnamefont {A.~H.}\ \bibnamefont
			{MacDonald}},\ }\href@noop {} {\bibfield  {journal} {\bibinfo  {journal}
			{Physical Review B}\ }\textbf {\bibinfo {volume} {97}},\ \bibinfo {pages}
		{201301(R)} (\bibinfo {year} {2018})}\BibitemShut {NoStop}%
    \bibitem [{\citenamefont {Shi}\ \emph {et~al.}(2006)\citenamefont {Shi},
		\citenamefont {Zhang}, \citenamefont {Xiao},\ and\ \citenamefont
		{Niu}}]{shi2006}%
	\BibitemOpen
	\bibfield  {author} {\bibinfo {author} {\bibfnamefont {J.}~\bibnamefont
			{Shi}}, \bibinfo {author} {\bibfnamefont {P.}~\bibnamefont {Zhang}}, \bibinfo
		{author} {\bibfnamefont {D.}~\bibnamefont {Xiao}}, \ and\ \bibinfo {author}
		{\bibfnamefont {Q.}~\bibnamefont {Niu}},\ }\href@noop {} {\bibfield
		{journal} {\bibinfo  {journal} {Physical review letters}\ }\textbf {\bibinfo
			{volume} {96}},\ \bibinfo {pages} {076604} (\bibinfo {year}
		{2006})}\BibitemShut {NoStop}%
    \bibitem [{\citenamefont {Tatara}\ and\ \citenamefont {Kohno}(2004)}]{Tatara2004}%
    \BibitemOpen
	\bibfield  {author} {\bibinfo {author} {\bibfnamefont {G.}~\bibnamefont
			{Tatara}}\ and\ \bibinfo {author} {\bibfnamefont {H.}\ \bibnamefont {Kohno}},\
	}\href@noop {} {\bibfield  {journal} {\bibinfo  {journal} {Physical Review
				Letters}\ }\textbf {\bibinfo {volume} {92}},\ \bibinfo {pages} {086601} (\bibinfo
		{year} {2004})}\BibitemShut {NoStop}%
    \bibitem [{\citenamefont {Tserkovnyak}\ , \ \citenamefont {Skadsem}\ and\ \citenamefont
			{Brataas}(2006)}]{tserkovnyak2006}%
    \BibitemOpen
	\bibfield  {author} {\bibinfo {author} {\bibfnamefont {Y. }~\bibnamefont
			{Tserkovnyak}}\ , \ \bibinfo {author} {\bibfnamefont {H.~J. }\ \bibnamefont {Skadsem}}\ and\ \bibinfo {author} {\bibfnamefont {A. }~\bibnamefont
			{Brataas}},\
	}\href@noop {} {\bibfield  {journal} {\bibinfo  {journal} {Physical Review
				B}\ }\textbf {\bibinfo {volume} {74}},\ \bibinfo {pages} {144405} (\bibinfo
		{year} {2006})}\BibitemShut {NoStop}%
    \bibitem [{\citenamefont {Duine}\ , \ \citenamefont {N\'u\~nez}\ , \ \citenamefont
            {Sinova}\ and\ \citenamefont {Macdonald}(2007)}]{Duine2007}%
    \BibitemOpen
	\bibfield  {author} {\bibinfo {author} {\bibfnamefont {R.~A. }~\bibnamefont
			{Duine}}\ , \ \bibinfo {author} {\bibfnamefont {A.~S. }\ \bibnamefont {N\'u\~nez}}\ , \ \bibinfo {author} {\bibfnamefont {J. }~\bibnamefont
			{Sinova}}\ and\ \bibinfo {author} {\bibfnamefont {A.~H. }~\bibnamefont
			{Macdonald}},\
	}\href@noop {} {\bibfield  {journal} {\bibinfo  {journal} {Physical Review
				B}\ }\textbf {\bibinfo {volume} {75}},\ \bibinfo {pages} {214420} (\bibinfo
		{year} {2007})}\BibitemShut {NoStop}%
    \bibitem [{\citenamefont {Hals}\ and\ \citenamefont
		{Brataas}(2015)}]{Hals2015}%
	\BibitemOpen
	\bibfield  {author} {\bibinfo {author} {\bibfnamefont {K.~M.~D.}\
			\bibnamefont {Hals}}\ and\ \bibinfo {author} {\bibfnamefont {A.}~\bibnamefont
			{Brataas}},\ }\href {\doibase 10.1103/PhysRevB.91.214401} {\bibfield
		{journal} {\bibinfo  {journal} {Physical Review B}\ }\textbf {\bibinfo
			{volume} {91}},\ \bibinfo {pages} {214401} (\bibinfo {year}
		{2015})}\BibitemShut {NoStop}%
	\bibitem [{\citenamefont {Hals}\ and\ \citenamefont
		{Brataas}(2013)}]{hals2013}%
	\BibitemOpen
	\bibfield  {author} {\bibinfo {author} {\bibfnamefont {K.~M.~D.}\ \bibnamefont
			{Hals}}\ and\ \bibinfo {author} {\bibfnamefont {A.}~\bibnamefont {Brataas}},\
	}\href@noop {} {\bibfield  {journal} {\bibinfo  {journal} {Physical Review
				B}\ }\textbf {\bibinfo {volume} {88}},\ \bibinfo {pages} {085423} (\bibinfo
		{year} {2013})}\BibitemShut {NoStop}%
	\bibitem [{\citenamefont {Liu}\ \emph {et~al.}(2010)\citenamefont {Liu},
		\citenamefont {Qi}, \citenamefont {Zhang}, \citenamefont {Dai}, \citenamefont
		{Fang},\ and\ \citenamefont {Zhang}}]{Liu2010}%
	\BibitemOpen
	\bibfield  {author} {\bibinfo {author} {\bibfnamefont {C.-X.}\ \bibnamefont
			{Liu}}, \bibinfo {author} {\bibfnamefont {X.-L.}\ \bibnamefont {Qi}},
		\bibinfo {author} {\bibfnamefont {H.~J.}~\bibnamefont {Zhang}}, \bibinfo
		{author} {\bibfnamefont {X.}~\bibnamefont {Dai}}, \bibinfo {author}
		{\bibfnamefont {Z.}~\bibnamefont {Fang}}, \ and\ \bibinfo {author}
		{\bibfnamefont {S.-C.}\ \bibnamefont {Zhang}},\ }\href {\doibase
		10.1103/PhysRevB.82.045122} {\bibfield  {journal} {\bibinfo  {journal} {Phys.
				Rev. B}\ }\textbf {\bibinfo {volume} {82}},\ \bibinfo {pages} {045122}
		(\bibinfo {year} {2010})}\BibitemShut {NoStop}%
	\bibitem [{\citenamefont {Schrieffer}\ and\ \citenamefont
		{Wolff}(1966)}]{Schrieffer1966}%
	\BibitemOpen
	\bibfield  {author} {\bibinfo {author} {\bibfnamefont {J.~R.}\ \bibnamefont
			{Schrieffer}}\ and\ \bibinfo {author} {\bibfnamefont {P.~A.}\ \bibnamefont
			{Wolff}},\ }\href {\doibase 10.1103/PhysRev.149.491} {\bibfield  {journal}
		{\bibinfo  {journal} {Physical Review}\ }\textbf {\bibinfo {volume} {149}},\
		\bibinfo {pages} {491} (\bibinfo {year} {1966})}\BibitemShut {NoStop}%
	\bibitem [{\citenamefont {Winkler}(2003)}]{Winkler2003}%
	\BibitemOpen
	\bibfield  {author} {\bibinfo {author} {\bibfnamefont {R.}~\bibnamefont
			{Winkler}},\ }\href {\doibase 10.1007/b13586} {\emph {\bibinfo {title}
			{Spin--Orbit Coupling Effects in Two-Dimensional Electron and Hole
				Systems}}}\ (\bibinfo {year} {2003})\BibitemShut {NoStop}%
    \bibitem [{\citenamefont {Akzyanov}\ and\ \citenamefont
		{Rakhmanov}(2018)}]{Akzyanov20182}%
	\BibitemOpen
	\bibfield  {author} {\bibinfo {author} {\bibfnamefont {R.~S.}\ \bibnamefont
			{Akzyanov}}\ and\ \bibinfo {author} {\bibfnamefont {A.~L.}\ \bibnamefont
			{Rakhmanov}},\ }\href {\doibase 10.1103/PhysRevB.99.045436} {\bibfield
		{journal} {\bibinfo  {journal} {Physical Review B}\ }\textbf {\bibinfo
			{volume} {99}},\ \bibinfo {pages} {045436} (\bibinfo {year}
		{2019})}\BibitemShut {NoStop}%
	\bibitem [{\citenamefont {Liu}\ \emph {et~al.}(2015)\citenamefont {Liu},
		\citenamefont {Richardella}, \citenamefont {Garate}, \citenamefont {Zhu},
		\citenamefont {Samarth},\ and\ \citenamefont {Chen}}]{liu2015}%
	\BibitemOpen
	\bibfield  {author} {\bibinfo {author} {\bibfnamefont {L.}~\bibnamefont
			{Liu}}, \bibinfo {author} {\bibfnamefont {A.}~\bibnamefont {Richardella}},
		\bibinfo {author} {\bibfnamefont {I.}~\bibnamefont {Garate}}, \bibinfo
		{author} {\bibfnamefont {Y.}~\bibnamefont {Zhu}}, \bibinfo {author}
		{\bibfnamefont {N.}~\bibnamefont {Samarth}}, \ and\ \bibinfo {author}
		{\bibfnamefont {C.-T.}\ \bibnamefont {Chen}},\ }\href@noop {} {\bibfield
		{journal} {\bibinfo  {journal} {Physical Review B}\ }\textbf {\bibinfo
			{volume} {91}},\ \bibinfo {pages} {235437} (\bibinfo {year}
		{2015})}\BibitemShut {NoStop}%
    \bibitem [{\citenamefont {Wigner}(1932)}]{Wigner1932}%
	\BibitemOpen
	\bibfield  {author} {\bibinfo {author} {\bibfnamefont {E.}~\bibnamefont
			{Wigner}},\ }\href@noop {} {\bibfield  {journal} {\bibinfo  {journal}
			{Physical Review}\ }\textbf {\bibinfo {volume} {40}},\ \bibinfo {pages} {749}
		(\bibinfo {year} {1932})}\BibitemShut {NoStop}%
    \bibitem [{\citenamefont {Sinitsyn},\citenamefont {MacDonald}, \citenamefont {Jungwirth}, \citenamefont {Dugaev}\ and\ \citenamefont {Sinova}(2007)}]{Sinitsyn2007}%
	\BibitemOpen
	\bibfield  {author} {\bibinfo {author} {\bibfnamefont {N.~A.}~\bibnamefont
			{Sinitsyn}}, \bibinfo {author} {\bibfnamefont {A.~H. }~\bibnamefont {MacDonald}}, \bibinfo
		  {author} {\bibfnamefont {T. }~\bibnamefont {Jungwirth}}, \bibinfo {author}  {\bibfnamefont {V.~K. }~\bibnamefont {Dugaev}} \ and\ \bibinfo {author} 
            {\bibfnamefont
			{J. }~\bibnamefont {Sinova}},\ }\href@noop {} {\bibfield  {journal} {\bibinfo
			{journal} {Physical Review B}\ }\textbf {\bibinfo {volume} {75}},\
		\bibinfo {pages} {045315} (\bibinfo {year} {2007})}\BibitemShut {NoStop}%
    \bibitem [{\citenamefont {Atencia},\citenamefont {Niu}\ and\ \citenamefont {Culcer}(2022)}]{Atencia2022}%
	\BibitemOpen
	\bibfield  {author} {\bibinfo {author} {\bibfnamefont {R.~B.}~\bibnamefont
			{Atencia}}, \bibinfo {author} {\bibfnamefont {Q. }~\bibnamefont {Niu}}\ and\ \bibinfo
		  {author} {\bibfnamefont {D. }~\bibnamefont {Culcer}},\ }\href@noop {} {\bibfield  {journal} {\bibinfo
			{journal} {Physical Review Research}\ }\textbf {\bibinfo {volume} {4}},\
		\bibinfo {pages} {013001} (\bibinfo {year} {2022})}\BibitemShut {NoStop}%
	\bibitem [{\citenamefont {Wang}\ \emph {et~al.}(2010)\citenamefont {Wang},
		\citenamefont {Lin}, \citenamefont {Wei}, \citenamefont {Liu}, \citenamefont
		{Dumas}, \citenamefont {Liu},\ and\ \citenamefont {Shi}}]{wang2010}%
	\BibitemOpen
	\bibfield  {author} {\bibinfo {author} {\bibfnamefont {Z.}~\bibnamefont
			{Wang}}, \bibinfo {author} {\bibfnamefont {T.}~\bibnamefont {Lin}}, \bibinfo
		{author} {\bibfnamefont {P.}~\bibnamefont {Wei}}, \bibinfo {author}
		{\bibfnamefont {X.}~\bibnamefont {Liu}}, \bibinfo {author} {\bibfnamefont
			{R.}~\bibnamefont {Dumas}}, \bibinfo {author} {\bibfnamefont
			{K.}~\bibnamefont {Liu}}, \ and\ \bibinfo {author} {\bibfnamefont
			{J.}~\bibnamefont {Shi}},\ }\href@noop {} {\bibfield  {journal} {\bibinfo
			{journal} {Applied physics letters}\ }\textbf {\bibinfo {volume} {97}},\
		\bibinfo {pages} {042112} (\bibinfo {year} {2010})}\BibitemShut {NoStop}%
	\bibitem [{\citenamefont {Choi}\ \emph {et~al.}(2012)\citenamefont {Choi},
		\citenamefont {Jo}, \citenamefont {Lee}, \citenamefont {Lee}, \citenamefont
		{Jo}, \citenamefont {Kajino}, \citenamefont {Takabatake}, \citenamefont {Ko},
		\citenamefont {Park},\ and\ \citenamefont {Jung}}]{choi2012}%
	\BibitemOpen
	\bibfield  {author} {\bibinfo {author} {\bibfnamefont {Y.}~\bibnamefont
			{Choi}}, \bibinfo {author} {\bibfnamefont {N.}~\bibnamefont {Jo}}, \bibinfo
		{author} {\bibfnamefont {K.}~\bibnamefont {Lee}}, \bibinfo {author}
		{\bibfnamefont {H.}~\bibnamefont {Lee}}, \bibinfo {author} {\bibfnamefont
			{Y.}~\bibnamefont {Jo}}, \bibinfo {author} {\bibfnamefont {J.}~\bibnamefont
			{Kajino}}, \bibinfo {author} {\bibfnamefont {T.}~\bibnamefont {Takabatake}},
		\bibinfo {author} {\bibfnamefont {K.-T.}\ \bibnamefont {Ko}}, \bibinfo
		{author} {\bibfnamefont {J.-H.}\ \bibnamefont {Park}}, \ and\ \bibinfo
		{author} {\bibfnamefont {M.}~\bibnamefont {Jung}},\ }\href@noop {} {\bibfield
		{journal} {\bibinfo  {journal} {Applied Physics Letters}\ }\textbf {\bibinfo
			{volume} {101}},\ \bibinfo {pages} {152103} (\bibinfo {year}
		{2012})}\BibitemShut {NoStop}%
    \bibitem [{\citenamefont {Gehring}\ \emph {et~al.}(2012)\citenamefont
		{Gehring}, \citenamefont {Gao}, \citenamefont {Burghard},\ and\ \citenamefont
		{Kern}}]{gehring2012}%
	\BibitemOpen
	\bibfield  {author} {\bibinfo {author} {\bibfnamefont {P.}~\bibnamefont
			{Gehring}}, \bibinfo {author} {\bibfnamefont {B.~F.}\ \bibnamefont {Gao}},
		\bibinfo {author} {\bibfnamefont {M.}~\bibnamefont {Burghard}}, \ and\
		\bibinfo {author} {\bibfnamefont {K.}~\bibnamefont {Kern}},\ }\href@noop {}
	{\bibfield  {journal} {\bibinfo  {journal} {Nano letters}\ }\textbf {\bibinfo
			{volume} {12}},\ \bibinfo {pages} {5137} (\bibinfo {year}
		{2012})}\BibitemShut {NoStop}%
	\bibitem [{\citenamefont {Wang}\ \emph {et~al.}(2013)\citenamefont {Wang},
		\citenamefont {Liu}, \citenamefont {Wang}, \citenamefont {Meyer},
		\citenamefont {Bao}, \citenamefont {He}, \citenamefont {Lang}, \citenamefont
		{Chen}, \citenamefont {Che}, \citenamefont {Post}, \citenamefont {Zou},
		\citenamefont {Basov}, \citenamefont {Wang},\ and\ \citenamefont
		{Xiu}}]{Wang2013}%
	\BibitemOpen
	\bibfield  {author} {\bibinfo {author} {\bibfnamefont {K.}~\bibnamefont
			{Wang}}, \bibinfo {author} {\bibfnamefont {Y.}~\bibnamefont {Liu}}, \bibinfo
		{author} {\bibfnamefont {W.}~\bibnamefont {Wang}}, \bibinfo {author}
		{\bibfnamefont {N.}~\bibnamefont {Meyer}}, \bibinfo {author} {\bibfnamefont
			{L.~H.}\ \bibnamefont {Bao}}, \bibinfo {author} {\bibfnamefont
			{L.}~\bibnamefont {He}}, \bibinfo {author} {\bibfnamefont {M.~R.}\
			\bibnamefont {Lang}}, \bibinfo {author} {\bibfnamefont {Z.~G.}\ \bibnamefont
			{Chen}}, \bibinfo {author} {\bibfnamefont {X.~Y.}\ \bibnamefont {Che}},
		\bibinfo {author} {\bibfnamefont {K.}~\bibnamefont {Post}}, \bibinfo {author}
		{\bibfnamefont {J.}~\bibnamefont {Zou}}, \bibinfo {author} {\bibfnamefont
			{D.~N.}\ \bibnamefont {Basov}}, \bibinfo {author} {\bibfnamefont {K.~L.}\
			\bibnamefont {Wang}}, \ and\ \bibinfo {author} {\bibfnamefont
			{F.}~\bibnamefont {Xiu}},\ }\href {\doibase 10.1063/1.4813903} {\bibfield
		{journal} {\bibinfo  {journal} {Applied Physics Letters}\ }\textbf {\bibinfo
			{volume} {103}},\ \bibinfo {pages} {031605} (\bibinfo {year} {2013})},\
	\Eprint {http://arxiv.org/abs/https://doi.org/10.1063/1.4813903}
	{https://doi.org/10.1063/1.4813903} \BibitemShut {NoStop}%
    \bibitem [{\citenamefont {Yan}\ and\ \citenamefont
          {Bazaliy}(2015)}]{yan2015phase}%
      \BibitemOpen
          \bibfield  {author} {\bibinfo {author} {\bibfnamefont {S.}~\bibnamefont
          {Yan}}\ and\ \bibinfo {author} {\bibfnamefont {Y.~B.}\ \bibnamefont
          {Bazaliy}},\ }\href@noop {} {\bibfield  {journal} {\bibinfo  {journal} {Physical Review
          B}\ }\textbf {\bibinfo {volume} {91}},\ \bibinfo {pages} {214424} (\bibinfo
          {year} {2015})}\BibitemShut {NoStop}%
    \bibitem [{\citenamefont {Liu}\ \emph {et~al.}(2012)\citenamefont {Liu},
          \citenamefont {Lee}, \citenamefont {Gudmundsen}, \citenamefont {Ralph},\ and\
          \citenamefont {Buhrman}}]{liu2012current}%
      \BibitemOpen
          \bibfield  {author} {\bibinfo {author} {\bibfnamefont {L.}~\bibnamefont
          {Liu}}, \bibinfo {author} {\bibfnamefont {O.}~\bibnamefont {Lee}}, \bibinfo
          {author} {\bibfnamefont {T.}~\bibnamefont {Gudmundsen}}, \bibinfo {author}
          {\bibfnamefont {D.}~\bibnamefont {Ralph}},\ and\ \bibinfo {author}
          {\bibfnamefont {R.}~\bibnamefont {Buhrman}},\ }\href@noop {}
          {\bibfield  {journal} {\bibinfo  {journal} {Physical review letters}\
          }\textbf {\bibinfo {volume} {109}},\ \bibinfo {pages} {096602} (\bibinfo
          {year} {2012})}\BibitemShut {NoStop}%
    \bibitem [{\citenamefont {Perez}\ \emph {et~al.}(2014)\citenamefont {Perez},
          \citenamefont {Martinez}, \citenamefont {Torres}, \citenamefont {Woo},
          \citenamefont {Emori},\ and\ \citenamefont {Beach}}]{perez2014chiral}%
      \BibitemOpen
          \bibfield  {author} {\bibinfo {author} {\bibfnamefont {N.}~\bibnamefont
          {Perez}}, \bibinfo {author} {\bibfnamefont {E.}~\bibnamefont {Martinez}},
          \bibinfo {author} {\bibfnamefont {L.}~\bibnamefont {Torres}}, \bibinfo
          {author} {\bibfnamefont {S.-H.}\ \bibnamefont {Woo}}, \bibinfo {author}
          {\bibfnamefont {S.}~\bibnamefont {Emori}},\ and\ \bibinfo {author}
          {\bibfnamefont {G.}~\bibnamefont {Beach}},\ }
          \href@noop {} {\bibfield  {journal} {\bibinfo  {journal} {Applied Physics
          Letters}\ }\textbf {\bibinfo {volume} {104}},\ \bibinfo {pages} {092403}
          (\bibinfo {year} {2014})}\BibitemShut {NoStop}%
    \bibitem [{\citenamefont {Mikuszeit}\ \emph {et~al.}(2015)\citenamefont
          {Mikuszeit}, \citenamefont {Boulle}, \citenamefont {Miron}, \citenamefont
          {Garello}, \citenamefont {Gambardella}, \citenamefont {Gaudin},\ and\
          \citenamefont {Buda-Prejbeanu}}]{mikuszeit2015spin}%
      \BibitemOpen
          \bibfield  {author} {\bibinfo {author} {\bibfnamefont {N.}~\bibnamefont
          {Mikuszeit}}, \bibinfo {author} {\bibfnamefont {O.}~\bibnamefont {Boulle}},
          \bibinfo {author} {\bibfnamefont {I.~M.}\ \bibnamefont {Miron}}, \bibinfo
          {author} {\bibfnamefont {K.}~\bibnamefont {Garello}}, \bibinfo {author}
          {\bibfnamefont {P.}~\bibnamefont {Gambardella}}, \bibinfo {author}
          {\bibfnamefont {G.}~\bibnamefont {Gaudin}},\ and\ \bibinfo {author}
          {\bibfnamefont {L.~D.}\ \bibnamefont {Buda-Prejbeanu}},\ }
          \href@noop {} {\bibfield  {journal} {\bibinfo
          {journal} {Physical Review B}\ }\textbf {\bibinfo {volume} {92}},\ \bibinfo
          {pages} {144424} (\bibinfo {year} {2015})}\BibitemShut {NoStop}%
    \bibitem [{\citenamefont {Legrand}\ \emph {et~al.}(2015)\citenamefont
          {Legrand}, \citenamefont {Ramaswamy}, \citenamefont {Mishra},\ and\
          \citenamefont {Yang}}]{legrand2015coherent}%
      \BibitemOpen
          \bibfield  {author} {\bibinfo {author} {\bibfnamefont {W.}~\bibnamefont
          {Legrand}}, \bibinfo {author} {\bibfnamefont {R.}~\bibnamefont {Ramaswamy}},
          \bibinfo {author} {\bibfnamefont {R.}~\bibnamefont {Mishra}},\ and\ \bibinfo
          {author} {\bibfnamefont {H.}~\bibnamefont {Yang}},\ }\href@noop {} {\bibfield  {journal} {\bibinfo  {journal} {Physical
          Review Applied}\ }\textbf {\bibinfo {volume} {3}},\ \bibinfo {pages} {064012}
          (\bibinfo {year} {2015})}\BibitemShut {NoStop}%
    \bibitem [{\citenamefont {Yu}\ \emph {et~al.}(2016)\citenamefont {Yu},
          \citenamefont {Qiu}, \citenamefont {Wu}, \citenamefont {Yoon}, \citenamefont
          {Deorani}, \citenamefont {Besbas}, \citenamefont {Manchon},\ and\
          \citenamefont {Yang}}]{yu2016spin}%
      \BibitemOpen
          \bibfield  {author} {\bibinfo {author} {\bibfnamefont {J.}~\bibnamefont
          {Yu}}, \bibinfo {author} {\bibfnamefont {X.}~\bibnamefont {Qiu}}, \bibinfo
          {author} {\bibfnamefont {Y.}~\bibnamefont {Wu}}, \bibinfo {author}
          {\bibfnamefont {J.}~\bibnamefont {Yoon}}, \bibinfo {author} {\bibfnamefont
          {P.}~\bibnamefont {Deorani}}, \bibinfo {author} {\bibfnamefont {J.~M.}\
          \bibnamefont {Besbas}}, \bibinfo {author} {\bibfnamefont {A.}~\bibnamefont
          {Manchon}},\ and\ \bibinfo {author} {\bibfnamefont {H.}~\bibnamefont
          {Yang}},\ }\href@noop {} {\bibfield  {journal} {\bibinfo  {journal} {Scientific
          reports}\ }\textbf {\bibinfo {volume} {6}},\ \bibinfo {pages} {32629}
          (\bibinfo {year} {2016})}\BibitemShut {NoStop}%
      \bibitem [{\citenamefont {Sakai}\ and\ \citenamefont
    		{Kohno}(2014)}]{sakai2014}%
    	\BibitemOpen
    	\bibfield  {author} {\bibinfo {author} {\bibfnamefont {A.}~\bibnamefont
    			{Sakai}}\ and\ \bibinfo {author} {\bibfnamefont {H.}~\bibnamefont {Kohno}},\
    	}\href@noop {} {\bibfield  {journal} {\bibinfo  {journal} {Physical Review
    				B}\ }\textbf {\bibinfo {volume} {89}},\ \bibinfo {pages} {165307} (\bibinfo
    		{year} {2014})}\BibitemShut {NoStop}%
    	\bibitem [{\citenamefont {Fischer}\ \emph {et~al.}(2016)\citenamefont
    		{Fischer}, \citenamefont {Vaezi}, \citenamefont {Manchon},\ and\
    		\citenamefont {Kim}}]{Fischer2016}%
    	\BibitemOpen
    	\bibfield  {author} {\bibinfo {author} {\bibfnamefont {M.~H.}\ \bibnamefont
    			{Fischer}}, \bibinfo {author} {\bibfnamefont {A.}~\bibnamefont {Vaezi}},
    		\bibinfo {author} {\bibfnamefont {A.}~\bibnamefont {Manchon}}, \ and\
    		\bibinfo {author} {\bibfnamefont {E.~A.}\ \bibnamefont {Kim}},\ }\href
    	{\doibase 10.1103/PHYSREVB.93.125303/FIGURES/4/MEDIUM} {\bibfield  {journal}
    		{\bibinfo  {journal} {Physical Review B}\ }\textbf {\bibinfo {volume} {93}},\
    		\bibinfo {pages} {125303} (\bibinfo {year} {2016})},\ \Eprint
    	{http://arxiv.org/abs/1305.1328} {arXiv:1305.1328} \BibitemShut {NoStop}%
    	\bibitem [{\citenamefont {Ndiaye}\ \emph {et~al.}(2017)\citenamefont {Ndiaye},
    		\citenamefont {Akosa}, \citenamefont {Fischer}, \citenamefont {Vaezi},
    		\citenamefont {Kim},\ and\ \citenamefont {Manchon}}]{Ndiaye2017}%
    	\BibitemOpen
	\bibfield  {author} {\bibinfo {author} {\bibfnamefont {P.~B.}\ \bibnamefont
			{Ndiaye}}, \bibinfo {author} {\bibfnamefont {C.~A.}\ \bibnamefont {Akosa}},
		\bibinfo {author} {\bibfnamefont {M.~H.}\ \bibnamefont {Fischer}}, \bibinfo
		{author} {\bibfnamefont {A.}~\bibnamefont {Vaezi}}, \bibinfo {author}
		{\bibfnamefont {E.~A.}\ \bibnamefont {Kim}}, \ and\ \bibinfo {author}
		{\bibfnamefont {A.}~\bibnamefont {Manchon}},\ }\href {\doibase
		10.1103/PHYSREVB.96.014408/FIGURES/2/MEDIUM} {\bibfield  {journal} {\bibinfo
			{journal} {Physical Review B}\ }\textbf {\bibinfo {volume} {96}},\ \bibinfo
		{pages} {014408} (\bibinfo {year} {2017})}\BibitemShut {NoStop}%
      \bibitem [{\citenamefont {Haney}\ \emph {et~al.}(2013)\citenamefont {Haney},
          \citenamefont {Lee}, \citenamefont {Lee}, \citenamefont {Manchon},\ and\
          \citenamefont {Stiles}}]{haney2013}%
      \BibitemOpen
          \bibfield  {author} {\bibinfo {author} {\bibfnamefont {P.~M.}\ \bibnamefont
          {Haney}}, \bibinfo {author} {\bibfnamefont {H.-W.}\ \bibnamefont {Lee}},
          \bibinfo {author} {\bibfnamefont {K.-J.}\ \bibnamefont {Lee}}, \bibinfo
          {author} {\bibfnamefont {A.}~\bibnamefont {Manchon}},\ and\ \bibinfo {author}
          {\bibfnamefont {M.~D.}\ \bibnamefont {Stiles}},\ }\href@noop {} {\bibfield  {journal}
          {\bibinfo  {journal} {Physical Review B}\ }\textbf {\bibinfo {volume} {87}},\
          \bibinfo {pages} {174411} (\bibinfo {year} {2013})}\BibitemShut {NoStop}%
    \bibitem [{\citenamefont {Manchon}(2012)}]{manchon2012}%
      \BibitemOpen
          \bibfield  {author} {\bibinfo {author} {\bibfnamefont {A.}~\bibnamefont
          {Manchon}},\ }\href@noop {} {\bibfield  {journal}
          {\bibinfo  {journal} {arXiv preprint arXiv:1204.4869}\ } (\bibinfo {year}
          {2012})}\BibitemShut {NoStop}%
    \bibitem [{\citenamefont {Kim}\ \emph {et~al.}(2012)\citenamefont {Kim},
          \citenamefont {Seo}, \citenamefont {Ryu}, \citenamefont {Lee},\ and\
          \citenamefont {Lee}}]{kim2012}%
      \BibitemOpen
          \bibfield  {author} {\bibinfo {author} {\bibfnamefont {K.-W.}\ \bibnamefont
          {Kim}}, \bibinfo {author} {\bibfnamefont {S.-M.}\ \bibnamefont {Seo}},
          \bibinfo {author} {\bibfnamefont {J.}~\bibnamefont {Ryu}}, \bibinfo {author}
          {\bibfnamefont {K.-J.}\ \bibnamefont {Lee}},\ and\ \bibinfo {author}
          {\bibfnamefont {H.-W.}\ \bibnamefont {Lee}},\ }\href@noop {}
          {\bibfield  {journal} {\bibinfo  {journal} {Physical Review B}\ }\textbf
          {\bibinfo {volume} {85}},\ \bibinfo {pages} {180404} (\bibinfo {year}
          {2012})}\BibitemShut {NoStop}%
    \bibitem [{\citenamefont {Wang}\ and\ \citenamefont
          {Manchon}(2012)}]{wang2012diff}%
      \BibitemOpen
          \bibfield  {author} {\bibinfo {author} {\bibfnamefont {X.}~\bibnamefont
          {Wang}}\ and\ \bibinfo {author} {\bibfnamefont {A.}~\bibnamefont {Manchon}},\
          } \href@noop {}
          {\bibfield  {journal} {\bibinfo  {journal} {Physical review letters}\
          }\textbf {\bibinfo {volume} {108}},\ \bibinfo {pages} {117201} (\bibinfo
          {year} {2012})}\BibitemShut {NoStop}%
    \bibitem [{\citenamefont {Sokolewicz}\ \emph {et~al.}(2019)\citenamefont
          {Sokolewicz}, \citenamefont {Ado}, \citenamefont {Katsnelson}, \citenamefont
          {Ostrovsky},\ and\ \citenamefont {Titov}}]{sokolewicz2019}%
      \BibitemOpen
          \bibfield  {author} {\bibinfo {author} {\bibfnamefont {R.}~\bibnamefont
          {Sokolewicz}}, \bibinfo {author} {\bibfnamefont {I.}~\bibnamefont {Ado}},
          \bibinfo {author} {\bibfnamefont {M.}~\bibnamefont {Katsnelson}}, \bibinfo
          {author} {\bibfnamefont {P.}~\bibnamefont {Ostrovsky}},\ and\ \bibinfo
          {author} {\bibfnamefont {M.}~\bibnamefont {Titov}},\ }\href@noop {} {\bibfield  {journal}
          {\bibinfo  {journal} {Physical Review B}\ }\textbf {\bibinfo {volume} {99}},\
          \bibinfo {pages} {214444} (\bibinfo {year} {2019})}\BibitemShut {NoStop}%
	\bibitem [{\citenamefont {Amin}\ \emph {et~al.}(2018)\citenamefont {Amin},
		\citenamefont {Zemen},\ and\ \citenamefont {Stiles}}]{amin2018}%
	\BibitemOpen
	\bibfield  {author} {\bibinfo {author} {\bibfnamefont {V.~P.}\ \bibnamefont
			{Amin}}, \bibinfo {author} {\bibfnamefont {J.}~\bibnamefont {Zemen}}, \ and\
		\bibinfo {author} {\bibfnamefont {M.~D.}\ \bibnamefont {Stiles}},\
	}\href@noop {} {\bibfield  {journal} {\bibinfo  {journal} {Physical review
				letters}\ }\textbf {\bibinfo {volume} {121}},\ \bibinfo {pages} {136805}
		(\bibinfo {year} {2018})}\BibitemShut {NoStop}%
	\bibitem [{\citenamefont {Li}\ \emph {et~al.}(2019{\natexlab{b}})\citenamefont
		{Li}, \citenamefont {Wang}, \citenamefont {Deng},\ and\ \citenamefont
		{Yang}}]{li20191}%
	\BibitemOpen
	\bibfield  {author} {\bibinfo {author} {\bibfnamefont {J.-Y.}\ \bibnamefont
			{Li}}, \bibinfo {author} {\bibfnamefont {R.-Q.}\ \bibnamefont {Wang}},
		\bibinfo {author} {\bibfnamefont {M.-X.}\ \bibnamefont {Deng}}, \ and\
		\bibinfo {author} {\bibfnamefont {M.}~\bibnamefont {Yang}},\ }\href@noop {}
	{\bibfield  {journal} {\bibinfo  {journal} {Physical Review B}\ }\textbf
		{\bibinfo {volume} {99}},\ \bibinfo {pages} {155139} (\bibinfo {year}
		{2019}{\natexlab{b}})}\BibitemShut {NoStop}%
	\bibitem [{\citenamefont {Yazyev}\ \emph {et~al.}(2010)\citenamefont {Yazyev},
		\citenamefont {Moore},\ and\ \citenamefont {Louie}}]{yazyev2010}%
	\BibitemOpen
	\bibfield  {author} {\bibinfo {author} {\bibfnamefont {O.~V.}\ \bibnamefont
			{Yazyev}}, \bibinfo {author} {\bibfnamefont {J.~E.}\ \bibnamefont {Moore}}, \
		and\ \bibinfo {author} {\bibfnamefont {S.~G.}\ \bibnamefont {Louie}},\
	}\href@noop {} {\bibfield  {journal} {\bibinfo  {journal} {Physical review
				letters}\ }\textbf {\bibinfo {volume} {105}},\ \bibinfo {pages} {266806}
		(\bibinfo {year} {2010})}\BibitemShut {NoStop}%
	\bibitem [{\citenamefont {Luo}\ and\ \citenamefont {Qi}(2013)}]{luo2013}%
	\BibitemOpen
	\bibfield  {author} {\bibinfo {author} {\bibfnamefont {W.}~\bibnamefont
			{Luo}}\ and\ \bibinfo {author} {\bibfnamefont {X.-L.}\ \bibnamefont {Qi}},\
	}\href@noop {} {\bibfield  {journal} {\bibinfo  {journal} {Physical Review
				B}\ }\textbf {\bibinfo {volume} {87}},\ \bibinfo {pages} {085431} (\bibinfo
		{year} {2013})}\BibitemShut {NoStop}%
\end{thebibliography}

%

\end{document}